# Substrate-Dependence of Monolayer $MoS_2$ Thermal Conductivity and Thermal Boundary Conductance

Alexander J. Gabourie[1], Çağıl Köroğlu[1], and Eric Pop[1,2,3,*]

[1]Department of Electrical Engineering, Stanford University, Stanford, CA 94305, USA

[2]Department of Materials Science & Engineering, Stanford University, Stanford, CA 94305, USA

[3]Precourt Institute for Energy, Stanford University, CA 94305, USA

*Contact: epop@stanford.edu

The thermal properties of two-dimensional (2D) materials, like $MoS_2$, are known to be affected by interactions with their environment, but this has primarily been studied only with $SiO_2$ substrates. Here, we compare the thermal conductivity (TC) and thermal boundary conductance (TBC) of monolayer $MoS_2$ on amorphous (a-) and crystalline (c-) $SiO_2$, AlN, $Al_2O_3$, and *h*-BN monolayers using molecular dynamics. The room temperature TC of $MoS_2$ is ~38 $Wm^{-1}K^{-1}$ on amorphous substrates and up to ~68 $Wm^{-1}K^{-1}$ on crystalline substrates, with most of the difference due to substrate interactions with long-wavelength $MoS_2$ phonons (< 2 THz). An *h*-BN monolayer used as a buffer between $MoS_2$ and the substrate causes the $MoS_2$ TC to increase by up to 50%. Length-dependent calculations reveal TC size effects below ~2 μm and show that the $MoS_2$ TC is size- but not substrate-limited below ~100 nm. We also find that the TBC of $MoS_2$ with c-$Al_2O_3$ is over twice that with c-AlN despite a similar $MoS_2$ TC on both, indicating that the TC and TBC could be tuned independently. Finally, we compare the thermal resistance of $MoS_2$ transistors on all substrates to show that $MoS_2$ TBC is the most important parameter for heat removal for long-channel (> 150 nm) devices, while TBC and TC are equally important for short channels. This work provides important insights for electro-thermal applications of 2D materials on various substrates.

## I. INTRODUCTION

Two-dimensional (2D) semiconductors are being investigated for three-dimensional integrated circuits (3D-ICs), opto-electronics, and flexible electronics [1-3], in large part due to their relatively good electron and hole mobility in a sub-nanometer thin material [4]. Thermal properties of 2D materials are also important in this context as the numerous interfaces and poor thermal conductivity (TC) substrates (e.g. $SiO_2$, polyimide) present in electronic applications lead to self-heating, which decreases transistor performance [5] and reliability [6] during operation [7]. The ultrathin nature of 2D materials is known to lead to a dependence of their electrical mobility [8] and TC [9] on so-called 'remote' phonons or impurities, i.e., those belonging to the materials below and above the 2D semiconductor. For example, our recent calculations [9] have shown that monolayer $MoS_2$ supported by amorphous $SiO_2$ suffers decreased TC compared to the freely suspended material, and that encasing the 2D material further lowers its TC.

However, the effects of insulators other than $SiO_2$ on $MoS_2$ thermal transport remain unknown as does the impact of insulator crystallinity. Moreover, the thermal boundary conductance (TBC) of the interface between $MoS_2$ and such insulators is not well understood, especially with respect to its dependence on substrate type. The TBC at 2D material van der Waals (vdW) interfaces could be a stronger bottleneck for heat dissipation in electronics because it is of the order ~15 $MWm^{-2}K^{-1}$ at room temperature [6,10,11], which is equivalent to the thermal resistance of nearly 100 nm of $SiO_2$ (i.e., the Kapitza length).

Here, we use atomistic molecular dynamics (MD) to calculate the TC and TBC of $MoS_2$ supported by $SiO_2$, AlN, and $Al_2O_3$, all technologically relevant insulators with a wide range of TCs themselves. We examine both amorphous and crystalline forms of the substrates to quantify their effect on the $MoS_2$ thermal properties. In addition, because electrical properties of 2D materials are known to improve with a hexagonal boron nitride (*h*-BN) interfacial layer [12,13], we also study the thermal consequences of adding an *h*-BN layer between $MoS_2$ and each substrate. The atomistic simulations enable a direct comparison of thermal properties between structures and provide insights into the frequency-dependence of TC, the phonon mean free path, and length-dependent TC. Finally, we develop an analytical model to evaluate the effectiveness of heat removal from back-gated $MoS_2$ transistors of various geometries and compare the impact of different substrates by using our calculated TCs and TBCs for supported $MoS_2$.

## II. METHODS

### A. SIMULATION MODELS

We use MD for all calculations of TC and TBC in this study. While the TC of crystalline materials can be described well by the Peierls-Boltzmann transport equation paired with density functional theory [14], these methods currently struggle to model systems with defects, interfaces, and amorphous materials [15]. In contrast, MD can calculate the TC of spatially complex structures without any modifications or approximations [16-18]. It also naturally incorporates all anharmonicities of a solid and does not make assumptions about the dynamics of the system [15,19]. This makes MD ideal for realistic TBC calculations, especially for supported 2D materials where concepts of cross-plane phonon group velocities are not well defined [20].

All results in this paper are calculated using the Graphics Processing Units Molecular Dynamics (GPUMD-v2.5.1) package [21-25]. The atomic interactions within and between materials is as follows: $MoS_2$ is modeled by the reactive empirical bond-order potential with a Lennard-Jones (LJ) addition (REBO-LJ) [26-28], which has been shown to predict thermal properties accurately [29]; $SiO_2$ is modeled by the Tersoff potential [30] parameterized by Munetoh *et al.* [31]; AlN [32] and $Al_2O_3$ [33] are modeled by the Vashishta potential, *h*-BN is modeled by the Tersoff potential parameterized by Sevik *et al.* [34], and all inter-material, vdW interactions are modeled by the LJ potential with Lorentz-Berthelot mixing rules (see section S1 of the supplement). Unless otherwise stated, the timestep for each simulation is 0.5 fs and the temperature is 300 K.

Figures 1(a-f) displays the $MoS_2$-with-substrate combinations considered in this work, including both amorphous and crystalline substrates. The structures with an *h*-BN monolayer inserted between the $MoS_2$ and substrate are shown in supplementary Fig. S1. For all simulations, we use a 4080 atom $MoS_2$ sheet with an area of $10.8 \times 11$ nm$^2$, which has been shown to be of sufficient size for TC calculations with periodic in-plane boundary conditions [9,29]. We fix the lateral dimensions of the simulation cell and construct all substrates to fit into those dimensions. This ensures that $MoS_2$ is strain-free regardless of the substrate, as a strain of 1% could change the TC by up to 5% [35]. We create supported $MoS_2$ structures by placing $MoS_2$ on each substrate (with or without *h*-BN) and minimizing the energy of each heterostructure to establish the correct vdW distances between materials. The monolayer thickness of $MoS_2$ is 6.15 Å [36] and *h*-BN is 3.33 Å [37]. The a-$SiO_2$, a-AlN, a-$Al_2O_3$, c-$SiO_2$ (quartz), c-AlN, and c-$Al_2O_3$ (sapphire) substrates are 2.7 nm, 3.5 nm, 2.9 nm, 2.7 nm, 2.95 nm, and 2.5 nm thick, respectively, as seen in Fig. 1. For details about structure creation, including information on the surface orientation of each substrate, please see section S2 of the supplement.

## B. COMPUTATIONAL METHODS

For TC calculations along the $MoS_2$, we use the homogeneous nonequilibrium MD (HNEMD) method [24,38] which has been shown to work well for supported 2D materials [9]. This method enables a direct calculation of TC, but, in practice, we calculate the running average of the TC with [24]

$$\kappa(t) = \frac{1}{t}\int_0^t \frac{\langle J(\tau)\rangle_{\mathrm{ne}}}{TVF_{\mathrm{e}}} d\tau , \tag{1}$$

where $T$ is the system temperature, $V$ is the system volume, $F_{\mathrm{e}}$ is the driving force parameter (simplified to a scalar due to the isotropic in-plane thermal conductivity of $MoS_2$ [29]), and $\langle J(\tau)\rangle_{\mathrm{ne}}$ is the non-equilibrium heat current due to the driving force. This non-equilibrium heat current can be spectrally decomposed using the spectral heat current (SHC) method [24,25,39,40]. The frequency-domain, spectral TC can be written as

$$\kappa(\omega) = \frac{2\tilde{K}(\omega)}{TVF_{\mathrm{e}}}, \tag{2}$$

where $\tilde{K}(\omega)$ is the Fourier transform of the virial-velocity correlation as defined in Refs. [24,25]. The SHC method is also more robust than other spectral TC methods when considering systems with more anharmonicity [25]. Note that the in-plane TC can be decomposed into contributions from in-plane atomic motion (dominant in longitudinal and transverse acoustic phonons) and out-of-plane atomic motion (dominant in flexural, or ZA, acoustic phonons) [23,24]. Simulation and procedural details of the HNEMD and SHC methods can be found in section S3 of the supplement.

The SHC method calculates the spectral TC if the structure has no temperature gradient and is large enough for diffusive thermal transport; however, it can also calculate the thermal *conductance* $G(\omega)$ [23,41] during NEMD simulations (which have large temperature gradients) with ballistic thermal transport. The details of our $G(\omega)$ calculations and NEMD simulations are in section S6 of the supplement. With both $\kappa(\omega)$ and $G(\omega)$, we also calculate size-related thermal properties such as the spectral phonon mean free path (MFP), $\lambda(\omega) = \kappa(\omega)/G(\omega)$ [24], and, subsequently, the length-dependent TC [24,41]

$$\kappa(L) = \int_0^\infty \frac{d\omega}{2\pi}\frac{\kappa(\omega)}{1+\lambda(\omega)/L}. \tag{3}$$

Here, $L$ is the length between the two thermal baths.

In addition to the in-plane TC, the cross-plane thermal transport may also change drastically when $MoS_2$ is supported by different substrates. To investigate this, we calculate the TBC between the $MoS_2$ sheet and its substrate using the approach to equilibrium MD (AEMD) technique [42-44]. In these simulations, we allow previously heated $MoS_2$ to cool into its substrate and we track the temperature difference ($\Delta T = T_{\mathrm{MoS_2}} - T_{\mathrm{sub}}$) between the two over time $t$. Modeling each structure as a resistive-capacitive ($RC$) thermal circuit, we extract the TBC $G$ using

$$\Delta T(t) = \Delta T_0 e^{-\left(\frac{1}{C_{\mathrm{MoS_2}}}+\frac{1}{C_{\mathrm{sub}}}\right)AGt}, \tag{4}$$

where $A$ is the contact area of the $MoS_2$ with its substrate, $\Delta T_0$ is the initial temperature difference, and $C$ is the heat capacity with subscripts denoting the respective material. Our AEMD setup and simulation details can be found in section S7 of the supplement.

## III. RESULTS AND DISCUSSION

Based on the models we have chosen for our MD simulations, there are three areas to consider when interpreting the TC and TBC of supported $MoS_2$: the vdW interactions, the mass differences, and the structure of each material. First, the vdW bonds, facilitated through the LJ potential, mediate any interaction between $MoS_2$ and a substrate. (Note that we are not considering other interfacial effects, like chemical bonding [45], but our materials form otherwise atomically intimate contacts.) The strength of vdW bonds significantly influences thermal properties of $MoS_2$; an increase in vdW strength increases TBC and decreases TC [9,44,46]. Our choice to use materials with elements next to each other on the periodic table (i.e., Al – Si, and N – O) results in LJ parameters that are similar across each of the substrates and, speaking to our second effect, similar elemental masses in each substrate. Thus, any significant changes in the TC of supported $MoS_2$ likely do not come from differences in the strength of the vdW bonds or atomic masses in this study.

This leaves the atomic structure as the primary differentiator when interpreting the TC and TBC of supported $MoS_2$ in this study. We know that amorphous materials have vastly different vibrational modes than their crystalline counterparts, resulting in vastly different interactions and phonon scattering events with $MoS_2$. We also know that thermal properties of different crystalline substrates can vary greatly too, with the TC of c-$SiO_2$ [47] being over an order of magnitude smaller than that of c-AlN [48,49]. The TCs of amorphous substrates tend to be similar, typically near 1 to 2 $Wm^{-1}K^{-1}$ [50-53]. Which differences and similarities in the substrates matter to the thermal properties of $MoS_2$ will be central to our discussion.

### A. Thermal Conductivity Results

#### 1. Total Thermal Conductivity

We first calculate the TC of freely suspended monolayer $MoS_2$ at room temperature, which serves as a reference value for all other supported $MoS_2$ structures. The TC of suspended $MoS_2$ is $122.0 \pm 1.45$ $Wm^{-1}K^{-1}$ with respective contributions from in-plane and out-of-plane atomic motion of $90.7 \pm 1.3$ $Wm^{-1}K^{-1}$ and $31.3 \pm 0.8$ $Wm^{-1}K^{-1}$. This value agrees well with recent suspended monolayer $MoS_2$ TC measurements which are around 100 $Wm^{-1}K^{-1}$ [11,54], and with previous calculations [9,29].

We then calculate the TC of $MoS_2$ supported by amorphous and crystalline $SiO_2$, AlN, and $Al_2O_3$, as well as those same structures with an *h*-BN interlayer between $MoS_2$ and the substrate. (See Fig. 1 and Fig. S1 of the supplement for structure visualizations, respectively.) The results of these calculations are shown in Fig. 1(g) and all TCs are listed in Table S2 of the supplement. We find that the TC of $MoS_2$ supported by any amorphous substrate is approximately the same, in the range of ~36 to 39 $Wm^{-1}K^{-1}$. However, the TC of $MoS_2$ supported by crystalline substrates can be notably different. When $MoS_2$ is on c-AlN or c-$Al_2O_3$, its TC is in the range of ~66 to 68 $Wm^{-1}K^{-1}$, whereas $MoS_2$ has a TC of only ~42 $Wm^{-1}K^{-1}$ when supported by c-$SiO_2$ (quartz). Of these structures, only $MoS_2$ on a-$SiO_2$ has been studied experimentally, with a reported $MoS_2$ TC of $63 \pm 22$ $Wm^{-1}K^{-1}$ [55], although a sputtered Ni capping layer is known to damage monolayer $MoS_2$ [45], likely affecting their TC measurement.

Next, we consider the effect of adding a single layer of *h*-BN between $MoS_2$ and each substrate. The gray, horizontal lines in Fig. 1(g) show the new TC of $MoS_2$ recalculated in these configurations. We find that adding an *h*-BN interlayer increases the TC of $MoS_2$ in all cases. The largest increase of TC—by ~20 $Wm^{-1}K^{-1}$—is seen in the systems on $SiO_2$ after the addition of the *h*-BN interlayer. For the other substrates, $MoS_2$ sees a larger increase in TC when *h*-BN is placed on amorphous substrates (by ~12 $Wm^{-1}K^{-1}$) compared to on crystalline substrates (by ~3.7 to 8.7 $Wm^{-1}K^{-1}$). Interestingly, the *h*-BN interlayer prevents the substrates from reducing the in-plane contributions to TC, which results in the larger total TCs of $MoS_2$. The out-of-plane contributions are approximately the same as without the *h*-BN interlayer or lower. Previous calculations suggested that the TC of graphene was higher when supported by an *h*-BN substrate than by an a-$SiO_2$ substrate [56], indicating that other supported 2D materials could also have higher TCs if interfaced with *h*-BN.

## 2. Frequency Dependence of Thermal Conductivity

By calculating the frequency-dependent (spectral) TC using the SHC method, we can better understand how each substrate affects $MoS_2$. We first calculate the room temperature spectral TC of freely suspended $MoS_2$, with the results shown in Fig. 2(a). Here, we only show the contributions from the acoustic modes (i.e., < 8 THz) as optical mode contributions are negligible. (See Fig. S4 of the supplement for more details.) We find that the TC of suspended $MoS_2$ is heavily influenced by low frequency phonons, with 50% of the total TC contributions from frequencies below ~2.1 THz. On the other hand, for substrate-supported $MoS_2$, we find that those crucial, low frequency contributions are the most greatly affected by the substrate, and the spectral TC contribution even decreases towards zero frequency as seen in Figs. 2(b)-(d). Part of this degradation is due to the various substrates almost completely suppressing TC contributions from out-of-plane atomic motion below ~2 THz (i.e., long-wavelength flexural modes; see Fig. S4), which is where ~50% of those contributions come from in suspended $MoS_2$.

From earlier discussion, we know that $MoS_2$ supported by the various amorphous substrates has a similar total TC. The dotted curves in Figs. 2(b)-(d) reveal that the frequency dependence of $MoS_2$ TC on each amorphous substrate is similar as well. [A direct comparison of these curves is in Fig. S7(a)]. To explain this, we turn to the vibrational density of states (VDOS) of each material, which represents the density of vibrational modes (or phonon modes, if in a crystal) at a given energy [57] and is shown in Fig. 3. The overlap of VDOS between two materials has been tied to the transmission of phonons at an interface in previous TBC studies [20,58-62] and can be used here to understand the interaction between the $MoS_2$ and its substrate, with the presumption that a smaller overlap of VDOS with $MoS_2$ will result in a smaller TC degradation in $MoS_2$.

Interestingly, despite the similar spectral TCs of $MoS_2$ supported by amorphous substrates, the VDOS of the amorphous substrates are different. We see that the VDOS of a-$SiO_2$ is larger at frequencies below 2 THz showing that, for the case of amorphous substrates, a larger overlap in VDOS with $MoS_2$ does not result in a lower $MoS_2$ TC. The similar TCs may partially be due to our choice of substrate materials as they have atomic species of similar mass, similar vdW bond strength to $MoS_2$, and, because all three are amorphous, similar atomic structures. Amorphous materials have a diverse set of vibrational modes consisting of propagons (sinusoidal, phonon-like modes), diffusons (delocalized, non-sinusoidal modes), and locons (localized vibrations) [63,64] which can possesses enough momentum and energy variation in their propagating and localized vibrations [65,66] to maximize phonon scattering in $MoS_2$ (i.e., the remote phonon scattering is not substrate limited). From an interaction perspective, these amorphous substrates likely look identical to $MoS_2$ despite their VDOS differences. This position is confirmed by vdW force spectrum calculations [46] in Figs. S10(a,c,e) which show that forces each amorphous substrate exert on $MoS_2$ are comparable.

In contrast, the spectral TC of $MoS_2$ on crystalline substrates, which is shown as solid lines in Figs. 2(b)-(d), is quite different amongst the materials. [A direct comparison of these spectral TC curves is in Fig. S7(b).] We find qualitative similarities between the spectral TC for $MoS_2$ on c-$SiO_2$ and c-AlN, with both having a peak spectral TC at ~3 THz; however, spectral TC contributions for $MoS_2$ on c-AlN are much larger, with its peak ~2× larger than for $MoS_2$ on c-$SiO_2$. For $MoS_2$ on c-$Al_2O_3$ (sapphire), we see that contributions to TC remain high, even down to 0 THz, similar to suspended $MoS_2$. This suggests lower-frequency, longer-wavelength phonons are not as severely affected when $MoS_2$ is on c-$Al_2O_3$ as the other substrates. This observation is in alignment with our vdW force spectrum calculations which show that c-$Al_2O_3$ exerts the smallest force on $MoS_2$ in the range of 0 to 4 THz compared to all the other substrates. A more detailed look at these calculations can be found in section S5 of the supplement.

Interestingly, the spectral TCs of $MoS_2$ on c-$SiO_2$ and a-$SiO_2$ [Fig. 2(b)] are found to be comparable, with total TCs less than ~4 $Wm^{-1}K^{-1}$ apart. The VDOS of both substrates over 0 to 8 THz (where acoustic modes of $MoS_2$ are located) are similar too [Fig. 3(a)], suggesting that crystalline substrates with a large VDOS overlap with $MoS_2$ can also provide sufficient interaction (i.e., vibrational modes with proper energies and momenta to scatter $MoS_2$ phonons) to significantly reduce the TC of supported $MoS_2$. In contrast, we find that the VDOS of c-AlN and c-$Al_2O_3$ are ~2 to 5 times smaller than of a-AlN and a-$Al_2O_3$ in the 0 to 4 THz range, respectively [see Figs. 3(b,c)]. The spectral TCs of c-AlN and c-$Al_2O_3$, in Figs. 2(c,d), differ the most from their amorphous counterparts within this frequency range which suggests that, with a smaller VDOS overlap, the remote phonon scattering from the c-AlN and c-$Al_2O_3$ substrates is limited compared to a-AlN and a-$Al_2O_3$. However, we note that, despite their similar VDOS in the 0 to 4 THz range [see Fig. S8(b)], c-AlN and c-$Al_2O_3$ yield very different spectral TCs, emphasizing that VDOS overlap alone cannot be used to fully characterize the changes in the TC of supported $MoS_2$.

Finally, we consider the effects of an *h*-BN interlayer on the spectral TC of $MoS_2$. Recall that the total TC of $MoS_2$ always increases when an *h*-BN interlayer is used, as shown in Fig. 1(g). The spectral TC reveals that the increase is entirely from contributions of phonon frequencies up to 3 THz (see Fig. S6). To some degree, the *h*-BN layer "blocks" the low-frequency interactions between $MoS_2$ and the substrate, and the shape of the *h*-BN-supported spectral TC curves become closer to resembling the suspended $MoS_2$ spectral TC curve in Fig. 2(a). Note that the contributions from out-of-plane motion, shown in Fig. S5, do not improve with an *h*-BN interlayer and that all TC gains are strictly from in-plane atomic motion (i.e., not flexural modes).

Since *h*-BN is a stiffer, high-TC material [55,67,68], it has, on average, 1.5 to 2 times fewer phonon modes than $MoS_2$ over the acoustic mode frequencies of $MoS_2$ (see Fig. S9), with this difference being as high as 7.7× between 0 THz to 2 THz. These observations suggest that remote phonon scattering with *h*-BN should be severely limited and the TC of $MoS_2$ should not be strongly affected by it. This is easily tested by calculating the TC of $MoS_2$ in a heterostructure with only *h*-BN, where we find that the TC of $MoS_2$ only drops 17% to ~101 $Wm^{-1}K^{-1}$ [see Fig. S5(a) and Table S2]. In structures with substrates, we expect the small number of *h*-BN phonon modes and the additional distance between $MoS_2$ and the substrate to limit remote phonon scattering in $MoS_2$.

### 3. Length Dependence of Thermal Conductivity

Next, we calculate the frequency-dependence of the phonon MFP as $\lambda(\omega) = \kappa(\omega)/G(\omega)$ [24], where the frequency-dependent conductance $G(\omega)$ describes ballistic thermal transport in $MoS_2$. The results of these calculations are shown in Figs. 4(a,b). If we examine the MFP of suspended $MoS_2$ in Fig. 4(a), we see that it drops rapidly from a peak of 3.3 μm at the Γ point (0 THz) to 10s of nanometers by 6 THz. Previous theoretical studies, which calculated a suspended $MoS_2$ TC similar to ours, have also shown MFPs up to several microns [69,70]. Other simulations which calculated peak MFPs in the 10s of nanometers, only reported TCs between 20 to 40 $Wm^{-1}K^{-1}$ [71-73]. Measurements of bulk in-plane TC also support claims of MFPs on the order of microns [74]. Note that it is the very long MFPs, in combination with high phonon group velocities at low phonon frequencies (i.e., ≤ 2 THz), that dominate the majority of in-plane thermal transport for suspended $MoS_2$, as shown in the spectral TC of Fig. 2(a).

When $MoS_2$ is supported by amorphous substrates, we find that the MFPs [also shown in Fig. 4(a)] are much smaller than in suspended $MoS_2$, with maximum MFPs in the range of a few hundred nanometers and significant deviations from suspended $MoS_2$ below 4 THz. This shows, again, that thermal transport and MFPs of long-wavelength phonons are severely disrupted by the amorphous substrates. In Fig. 4(b), we see that $MoS_2$ on c-$SiO_2$ (quartz) has a similar spectral MFP curve to a-$SiO_2$ in Fig. 4(a), and the c-AlN- and c-$Al_2O_3$-supported $MoS_2$ starts to see MFPs near 1 μm below 3 THz. Finally, if we consider the MFPs for systems with *h*-BN interlayers (Fig. S14), we find that the MFPs for systems of both amorphous and crystalline substrates increase significantly, with the MFP distribution more closely resembling that of suspended $MoS_2$.

With the spectral TC and MFP, we use Eq. (3) to calculate the length-dependent TC, as shown in Fig. 4(c). We find that the TC of suspended $MoS_2$ is length-dependent in samples shorter than ~10 μm and only converges to its bulk limit at longer length scales. Recent experimental work corroborates our results showing that the TC of $MoS_2$ continues to increase up to a suspended $MoS_2$ membrane diameter of ~13 μm [11]. In comparison, the TC of graphene, which is over an order of magnitude larger than that of $MoS_2$ [75,76], has been calculated to converge at length scales also an order of magnitude longer, around ~100 μm [77-79]. A special symmetry selection rule forbids certain phonon-phonon scattering events in graphene [76], enabling flexural modes with very long MFPs. As $MoS_2$ is three atoms thick, it does not follow this rule [80], resulting in diffusive thermal transport at a much shorter length than graphene.

The length-dependent trends for the TC of $MoS_2$ on other substrates [also shown in Fig. 4(c)] follow expectations based on their shorter MFPs. $MoS_2$ supported by crystalline substrates converges to its final value on the order of a few microns, and $MoS_2$ supported by amorphous substrates converges to its final value on the order of several hundred nanometers. There are some important implications of the length-dependent TC in samples shorter than ~100 nm. In this range, the TC of $MoS_2$ is size-limited no matter the substrate [including cases with *h*-BN interlayers, as seen in Fig. S14(c)]. This suggests some applications, such as patterned nanoscale transistors [7,81], may not see a significant TC benefit when choosing one substrate over another. In such devices, heat sinking may be dominated by the TBC with thermal pathways through the substrate, gate insulator, and gate (see part C of this section). However, the $MoS_2$-substrate vdW bond strength and force may still play a role, because a substrate that is more tightly coupled to $MoS_2$ could reduce the TC a non-negligible amount [9,46]. In addition, substrate-induced strain effects [82] could also influence the TC of $MoS_2$, although only of the order 5% [35].

## B. Thermal Boundary Conductance Results

Because the $MoS_2$-substrate TBC is a key property for heat removal in systems based on 2D materials, we run AEMD simulations and use Eq. (4) to extract the TBC for all structures near room temperature. These TBCs are shown with the TCs in Fig. 5 and additional details can be found in section S7 and Table S2 of the supplement. We find that the TBC for $MoS_2$ on a-$SiO_2$ and c-$SiO_2$ are similar, in the range of 22 to 23 $MWm^{-2}K^{-1}$. These TBCs are comparable to previous measurements and within the range of experimental error [6,10,11,83]. Based on the similar spectral TC, VDOS, and vdW force spectrum between the two $SiO_2$ substrates, this outcome is expected. Given that bulk c-$SiO_2$ (quartz) has a TC of 10.7 $Wm^{-1}K^{-1}$ parallel to the c-axis (or 6.7 $Wm^{-1}K^{-1}$ perpendicular to it) [47], whereas a-$SiO_2$ has a TC of ~1.4 $Wm^{-1}K^{-1}$ [84] near room temperature, c-$SiO_2$ is clearly preferred if trying to maximize heat removal in such circumstances.

Similarly, we find that the TBC of $MoS_2$ on c-$Al_2O_3$ is comparable to that on a-$Al_2O_3$, with both in the range of ~32 to 34 $MWm^{-2}K^{-1}$. This means that the sapphire substrate yields a high TBC and TC for $MoS_2$, making it superior for heat removal. This is especially true because the TC of bulk c-$Al_2O_3$ (~34 $Wm^{-1}K^{-1}$ [85]) is much higher than for a-$Al_2O_3$ (~1.6 $Wm^{-1}K^{-1}$ [53]) at room temperature. Understanding why the TBC of $MoS_2$ on c-$Al_2O_3$ is comparable to a-$Al_2O_3$ requires further investigation; however, we see in Fig. S10(f) that c-$Al_2O_3$ exerts a large vdW force on $MoS_2$ in the 4 to 8 THz range, which is not observed for other substrates. We hypothesize that this is responsible for the larger TBC at the $MoS_2$/c-$Al_2O_3$ interface. Note that this large vdW force also corresponds with a large VDOS overlap between c-$Al_2O_3$ and $MoS_2$ seen in Fig. 3(c) over the same frequency range. The a-$Al_2O_3$ substrate has a similar VDOS overlap but not a similar vdW force spectrum, suggesting that crystalline structure is the differentiator.

In contrast to the other substrates, the TBCs of $MoS_2$ on a-AlN and c-AlN are significantly different at ~29 $MWm^{-2}K^{-1}$ and ~14 $MWm^{-2}K^{-1}$, respectively. However, because the TC of a-AlN (~1.7 $Wm^{-1}K^{-1}$ [52]) is significantly lower than the TC of bulk c-AlN (>200 $Wm^{-1}K^{-1}$ [48]), it is unclear which substrate is best for heat removal, a scenario we will investigate in section III.C. While previous experimental work agrees with our TBC calculation of $MoS_2$ on c-AlN [10], we have difficulty explaining its smaller value compared to our TBC of $MoS_2$ on c-$Al_2O_3$. In Fig. 3(b), we also see a large VDOS overlap between $MoS_2$ and c-AlN in the 4 to 8 THz range, however, we do not see corresponding vdW forces like for $MoS_2$ on c-$Al_2O_3$ [see Figs. S10 (d),(f)]. This indicates that, for crystalline substrates, the VDOS overlap alone is insufficient to estimate the relative TBCs of supported $MoS_2$ and more detailed structural information must be considered. Future work leveraging techniques like interface conductance modal analysis (ICMA) [15,19] can provide the necessary structural and frequency-domain insights to fully understand these TBCs. Even without this analysis, the lack of a trend in Fig. 5 still shows that the interactions which affect the TC and TBC of substrate-supported $MoS_2$ are not the same (i.e., there is not necessarily a trade-off between the two properties).

Finally, we calculate the TBC of $MoS_2$ with each *h*-BN-capped substrate. These structures are thermally more complex, and we must consider two TBCs: one between $MoS_2$ and *h*-BN and another between *h*-BN and each substrate. Additional details about the thermal circuit and TBC extractions can be found in section S7 of the supplement. We find the limiting TBC to be at the $MoS_2$/*h*-BN interface, similar to experimental work on a-$SiO_2$ [86], with TBCs between 18 to 26 $MWm^{-2}K^{-1}$ across all substrates. Interestingly, when the *h*-BN layer is introduced, the vdW force on $MoS_2$ from c-$Al_2O_3$ in the 4 to 8 THz range vanishes [see Fig. S11(g)] and at the same time the TBC drops by nearly 2× (Table S2). This supports our hypothesis that the force from c-$Al_2O_3$ in that range of frequencies is responsible for the higher TBC between $MoS_2$ and c-$Al_2O_3$ (~34.3 $MWm^{-2}K^{-1}$ in Fig. 5 and Table S2).

## C. Transistor Modeling and Practical Considerations

To investigate the impact of the TC and TBC of supported $MoS_2$ on the temperature rise in a transistor due to self-heating, we have developed an analytical model for the thermal resistance of a simple back-gated $MoS_2$ transistor [7], shown in Fig. 6(a), and validated it through finite element method simulations (see section S9 of the supplement). Underneath the monolayer $MoS_2$ channel is a 50 nm-thick insulating film on a silicon substrate, which, as in many laboratory measurements, serves as the back gate. For consistency with our MD simulations, the insulating films will be limited to crystalline and amorphous $SiO_2$, AlN, and $Al_2O_3$. The $MoS_2$ lies between and underneath the two contacts, which are 100 nm thick and 500 nm long [i.e., *x*-direction in Fig. 6(a)]. In this section, we consider a wide device (10 µm in *y*-direction) to focus on the length dependence of the peak device temperature rather than on effects arising from narrow channel widths. (For narrow channel widths [87], the lateral heat spreading beyond $MoS_2$ in the *y*-direction, which increases the thermal footprint of the $MoS_2$ transistor, is non-negligible, and the $MoS_2$ thermal conductivity may be reduced by the narrow channel.) The peak thermal resistance is:

$$R_{th} = \frac{\Delta T_{max}}{P}, \tag{5}$$

where $P$ is the power assumed to be uniformly dissipated in the $MoS_2$ channel [6], consistent with a transistor operating in the linear regime, and $\Delta T_{max}$ is the resulting maximum temperature rise, which occurs at the center of the channel. The dependence of $R_{th}$ on the transistor geometry and material properties has previously been modeled as [87,88]:

$$R_{th} = \frac{1}{gL}\left[1-\left(\cosh\frac{L}{2L_H} + gL_H R_{con}\sinh\frac{L}{2L_H}\right)^{-1}\right], \tag{6}$$

where $L_H = (\kappa Wt/g)^{1/2}$ is the thermal healing length along the $MoS_2$, i.e., the characteristic distance over which the temperature drops by $1/e$ from the contacts. In addition, $g$ is the thermal conductance per unit length from the channel to the bottom of the Si substrate. Note that a substrate with a small TC like polyimide [2] can significantly reduce $g$, but, for the high-TC Si substrate ($\kappa_{sub}$ = 150 $Wm^{-1}K^{-1}$) considered in this section, the substrate contribution to $R_{th}$ is rather small, except for long, wide devices where the substrate can account for about half of $R_{th}$. The contact thermal resistance is $R_{con}$, and $L$, $W$, $t$, $\kappa$ are the $MoS_2$ channel length, width, thickness, and TC, respectively. The dimensions are labeled in Fig. 6(a). The $MoS_2$ TC affects $R_{th}$ through its influence on $R_{con}$ and $L_H$, while the $MoS_2$ TBC impacts $g$. The detailed descriptions of these quantities are given in section S9 of the supplement.

In the long channel limit ($L \gg 3L_H$), the effect of the contacts is negligible and $R_{th}$ depends purely on the heat transfer down into the supporting insulator and the Si substrate below it. In this case, we have $R_{th} \approx 1/(gL)$, which for a thin underlying insulator can be further approximated as

$$R_{th} \approx \frac{1}{WL \times \mathrm{TBC}_{\mathrm{MoS_2-ins}}} \tag{7}$$

where $W$ is the channel width and $\mathrm{TBC}_{\mathrm{MoS2\text{-}ins}}$ is the TBC between $MoS_2$ and the underlying insulator. (In this limit, when only TBC is considered, $3L_H \approx 150$ nm for all material interfaces in this work.) Equation (7) shows that the peak temperature is directly influenced by the $MoS_2$ TBC. If the thermal resistance contribution of the underlying insulator is not negligible, a better estimate is given by

$$R_{th} \approx \frac{\mathrm{TBC}^{-1}_{\mathrm{MoS_2\text{-}ins}} + \mathrm{TBC}^{-1}_{\mathrm{ins\text{-}sub}} + t_{\mathrm{ins}}/\kappa_{\mathrm{ins}}}{WL} \tag{8}$$

where $\mathrm{TBC}_{\mathrm{ins\text{-}sub}}$ is the TBC between the insulator and the Si substrate (which can usually be neglected when compared to other thermal resistances), $\kappa_{\mathrm{ins}}$ is the (vertical) insulator thermal conductivity and $t_{\mathrm{ins}}$ is the insulator thickness.

In the short channel limit ($L < 3L_H$), heat flow from $MoS_2$ directly into the insulator underneath is negligible, and heat removal is facilitated primarily by the contacts. In this limit, $R_{th} \approx R_{con}/2$, where the factor of two is the result of the two contacts providing heat sinking paths in parallel [89]. If the thermal resistance contributions of the underlying insulator and the contact metals are negligible, $R_{th}$ can be simplified as

$$R_{th} \approx \frac{1}{2W\sqrt{t\kappa\left(\mathrm{TBC}_{\mathrm{MoS_2-ins}} + \mathrm{TBC}_{\mathrm{MoS_2-met}}\right)}} \tag{9}$$

where $\mathrm{TBC}_{\mathrm{MoS2\text{-}met}}$ is the TBC from $MoS_2$ to the contact metals. Equation (9) depends only on the thermal properties of $MoS_2$ and its interfaces and can be viewed as the intrinsic thermal resistance of the $MoS_2$ channel, without contributions from external insulating layers. This also shows that the $MoS_2$ TC and total TBC to the materials on either side are equally important for short-channel devices. Note that $\mathrm{TBC}_{\mathrm{MoS2\text{-}met}}$ is typically on the same order of magnitude as $\mathrm{TBC}_{\mathrm{MoS2\text{-}ins}}$ and cannot be neglected. In this work, we assume $\mathrm{TBC}_{\mathrm{MoS2\text{-}met}} = 20$ $\mathrm{MWm^{-2}K^{-1}}$, which is typical of the interface with Au, Ti, and Al [74,90,91].

Equation (9) represents a good approximation for the transistor geometry and materials considered here as long as the contacts are longer than ~500 nm in the $x$-direction. In general, this expression will hold if the underlying insulator is thin (< ~100 nm) and the contact metals act as good heat sinks. Contacts are good heat sinks if they are moderately long (> ~500 nm), extend beyond the width of the device to connect to other devices, or join larger interconnects through vias. For shorter contacts, $R_{th}$ depends more on the TBC of $MoS_2$ than the TC for all channel lengths [see Figs. S18(b,d)]. In addition, $\mathrm{TBC}_{\mathrm{MoS2\text{-}met}}$ has a smaller impact on $R_{th}$ than $\mathrm{TBC}_{\mathrm{MoS2\text{-}ins}}$ does, because short metal contacts are of limited utility in cooling the device. In our simple model, the contacts are assumed to be exactly as wide as the channel and *not* connected to an external heat sink. They merely act as lateral heat spreaders, ultimately assisting heat flow into the substrate.

Figures 6(b,c) shows the dependence of $R_{th}$ on the transistor channel length $L$, for different underlying insulator materials. Note the distinct long-channel regime, where $R_{th}$ is inversely proportional to $L$ (except for very long channels where the substrate thermal resistance, which is proportional to $L^{-1/2}$, becomes significant), and the short channel regime, where $R_{th}$ is relatively independent of $L$. The channel length that roughly separates these two regimes is $3L_H \approx 150$ nm. For long channels, thanks to a higher TBC with $MoS_2$ and a relatively high cross-plane thermal conductivity (18 $\mathrm{Wm^{-1}K^{-1}}$) [92], the c-$Al_2O_3$ substrate achieves the lowest $R_{th}$. On the other hand, despite its high thermal conductivity (35 $\mathrm{Wm^{-1}K^{-1}}$) [48,93], the c-AlN

substrate results in a high $R_{th}$, exceeded only by a-$SiO_2$ due to its low TC (1.4 $Wm^{-1}K^{-1}$) [84], in accordance with Eq. (8). At short channels, consistent with Eq. (9), the c-$Al_2O_3$ substrate once again yields the lowest $R_{th}$ due to the high TC and TBC of $MoS_2$ when supported by it, followed by the c-AlN substrate which achieves the highest $MoS_2$ TC. Note that the TCs listed for the crystalline insulators account for size constraints (i.e., not bulk) and anisotropy (e.g., for quartz and sapphire).

We note that we have only considered back-gated $MoS_2$ transistors in this section, which is the simpler, more commonly encountered geometry in laboratory experiments. Top-gated transistors will be the subject of future work, as top gates have been found to play a more important role in heat spreading when the substrate TC is extremely low, such as is the case with polyimide [2].

## IV. CONCLUSIONS

We investigated the dependence of the TC of $MoS_2$ on crystalline and amorphous $SiO_2$, AlN, and $Al_2O_3$ substrates, showing that the TC of $MoS_2$ is larger when supported by crystalline substrates than amorphous and that amorphous substrates yield approximately the same TC for $MoS_2$ regardless of material. For any substrate, adding even a single *h*-BN interlayer improves the TC of $MoS_2$. The degradation of $MoS_2$ TC is primarily due to substrate interactions with long-wavelength, low-frequency $MoS_2$ phonons (i.e., < 2 THz), which a majority of suspended $MoS_2$ contributions arise from. The in-plane phonon mean free paths for $MoS_2$ can be very long (i.e., > 1 μm), and the diffusive thermal transport regime is not reached until length scales greater than a micron. Additionally, at lengths below 100 nm, the TC of $MoS_2$ becomes mostly limited by system size (i.e., $MoS_2$ length and width) and not substrate interactions. Finally, we found the TBC of $MoS_2$ to behave notably different for different crystalline substrates, i.e., the TBC with c-$Al_2O_3$ being over double that with c-AlN, despite yielding similar $MoS_2$ TCs. This suggests that the mechanisms controlling the TC and TBC of $MoS_2$ are different and that both could be optimized on a single substrate.

Finally, using an analytical model to determine the thermal resistance of a back-gated $MoS_2$ transistor, we evaluated the impact of our calculated TBCs and TCs of $MoS_2$ when supported by the insulating substrates considered in this study. Of these substrates, c-$Al_2O_3$ leads to the smallest temperature rise in an $MoS_2$ transistor. $MoS_2$ transistors on amorphous substrates will have larger temperature rises than on crystalline substrates as the TC of $MoS_2$ is worse and the low substrate TCs contribute significantly to the total device resistance. For transistors with long channels ($L > 3L_H \approx 150$ nm) or short contacts (< 500 nm) the TBC between $MoS_2$ and the underlying substrate is the dominant mechanism for heat removal. For short channel devices ($L < 3L_H \approx 150$ nm) with longer contacts (> 500 nm), the $MoS_2$ TC and total TBC with the contacts and substrate are equally important for heat removal. Overall, these TBC and TC results, as well as the analytical thermal model, provide important insights to evaluate heating effects during the operation of electronic and optical devices based on $MoS_2$ and similar 2D materials.

### Supplementary Material

See the supplementary material for details on our (S1) Lennard-Jones parameters, (S2) structure creation, (S3) homogeneous nonequilibrium simulations, (S4) vibrational density of states calculations, (S5) van der Waals force spectrum calculations, (S6) nonequilibrium simulations, (S7) approach to equilibrium calculations, (S8) final calculated thermal properties, and (S9) analytic thermal model.

### Acknowledgments

Some of the computing for this project was performed on the Sherlock cluster at Stanford University. We would like to thank Stanford University and the Stanford Research Computing Center (SRCC) for providing computational resources and support that contributed to these results. This work was also partially supported by the Stanford SystemX Alliance and by ASCENT, one of the six centers in JUMP, a Semiconductor Research Corporation (SRC) program sponsored by DARPA. A.J.G. also acknowledges support from the Achievement Rewards for College Scientists (ARCS) Northern California Chapter.

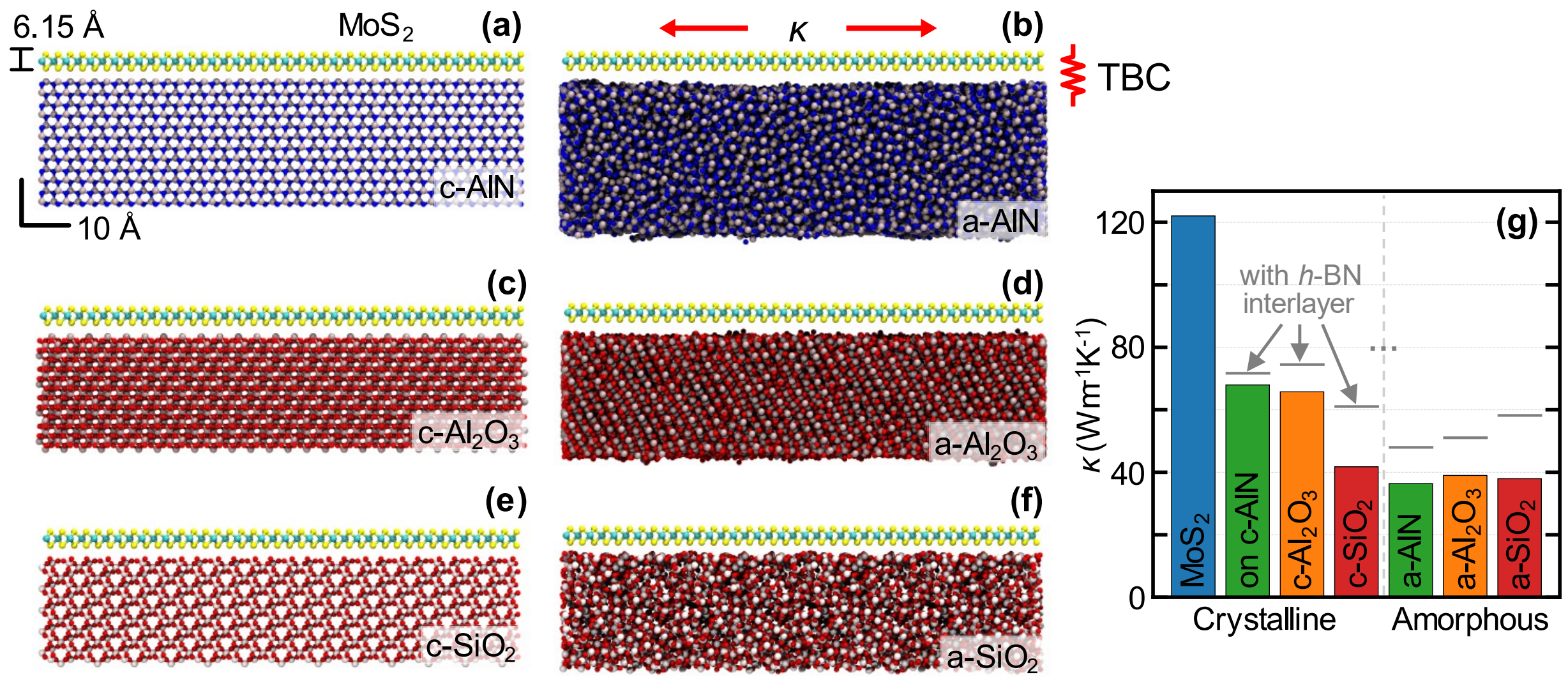

**Figure 1**: **(a)-(f)** Visualizations of the structures used for supported $MoS_2$ thermal conductivity and thermal boundary conductance calculations. The area of each structure is 10.8 × 11 nm$^2$ and the relative thicknesses can be determined from the scale bar in **(a)**. The in-plane and out-of-plane nature of thermal conductivity and thermal boundary conductance is highlighted by the red cartoons in **(b)**. **(g)** The thermal conductivity of monolayer $MoS_2$ when suspended (blue bar) and supported by crystalline (c-) and amorphous (a-) AlN, $Al_2O_3$, and $SiO_2$. The gray horizontal lines above the bars represents the thermal conductivity of $MoS_2$ if a single layer of *h*-BN is inserted between the $MoS_2$ and each substrate. All calculations are at 300 K.

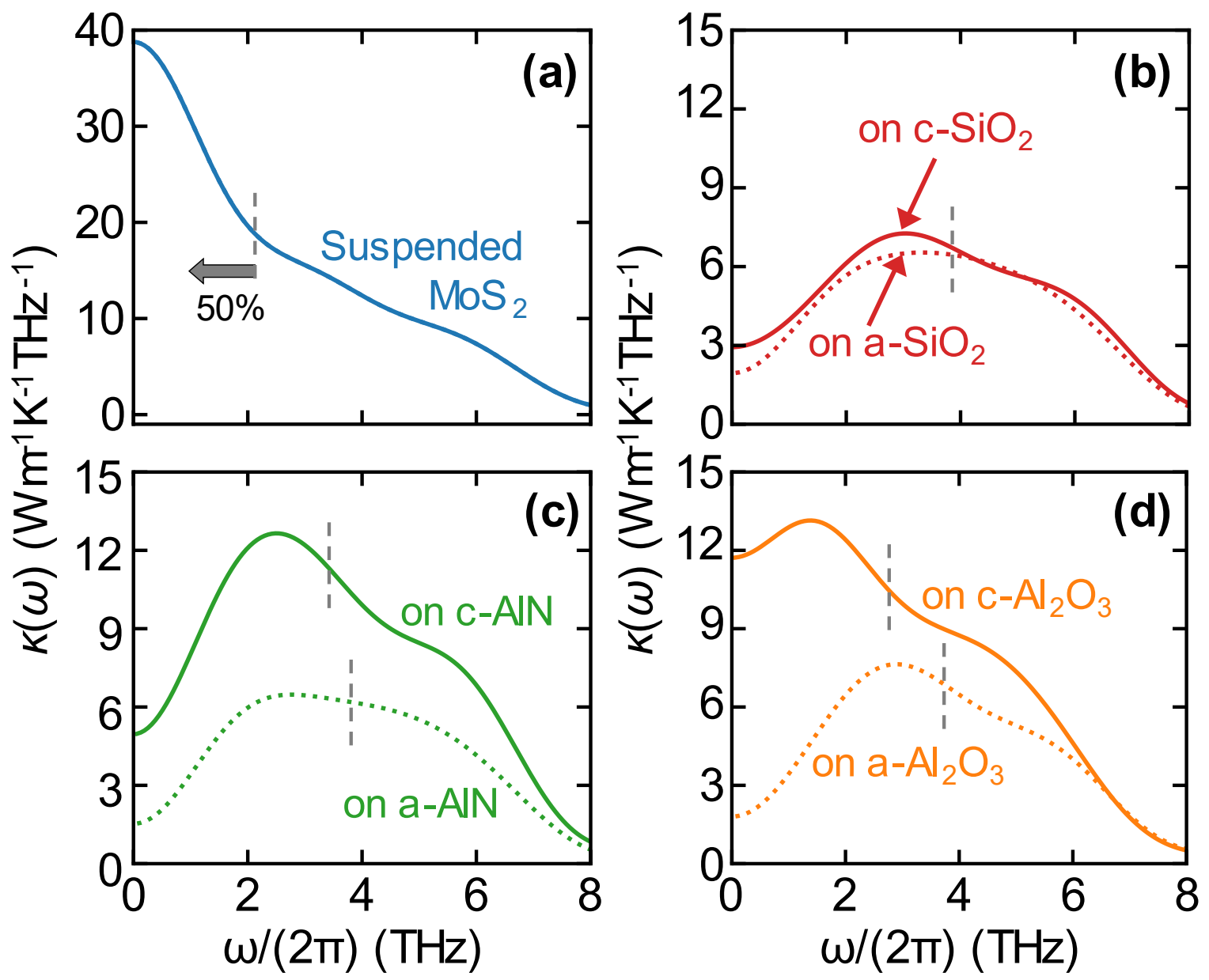

**Figure 2**: The frequency-dependent thermal conductivity $\kappa(\omega)$ when monolayer $MoS_2$ is **(a)** suspended or supported by **(b)** $SiO_2$, **(c)** AlN, and **(d)** $Al_2O_3$. The solid lines in **(b)-(d)** are for crystalline substrates and the dotted lines are for amorphous substrates. The gray vertical dashed lines show the phonon frequency at which 50% of the total thermal conductivity as has been contributed. All calculations are at 300 K.

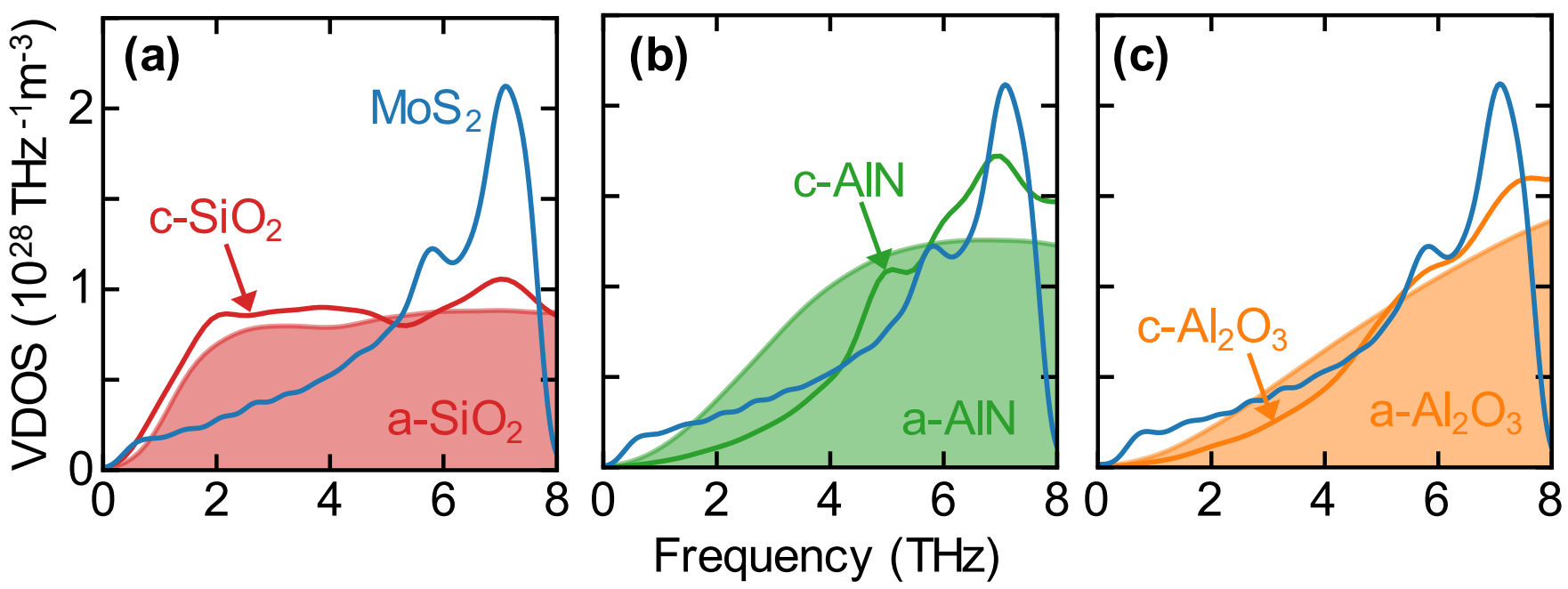

**Figure 3**: Vibrational density of states (VDOS) of crystalline (line) and amorphous (shaded area) **(a)** $SiO_2$, **(b)** AlN, and **(c)** $Al_2O_3$ compared to $MoS_2$ at 300 K. The $MoS_2$ curves are calculated when supported by the amorphous substrate corresponding to the panel.

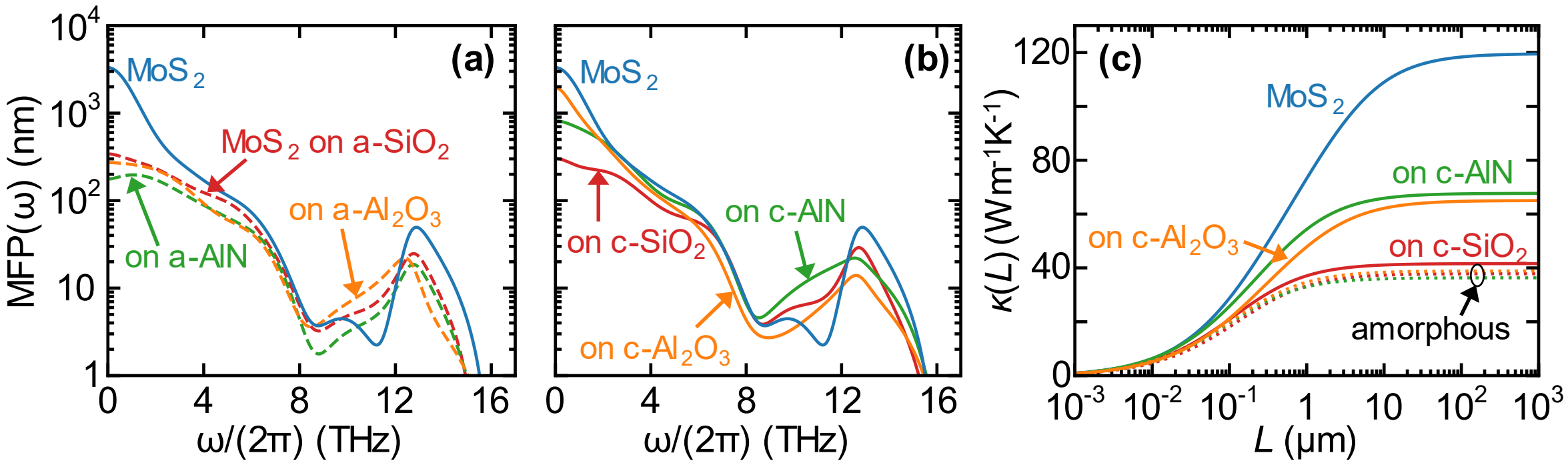

**Figure 4**: The frequency-dependent mean free path MFP($\omega$) of $MoS_2$ when supported by **(a)** amorphous substrates and **(b)** crystalline substrates at 300 K. Both **(a)** and **(b)** also show MFP($\omega$) for suspended $MoS_2$. **(c)** The room temperature length dependent thermal conductivity $\kappa(L)$ for suspended and supported $MoS_2$. The red, green, and orange dotted (solid) lines show $MoS_2$ when supported by amorphous (crystalline) substrates, and the blue line shows suspended $MoS_2$.

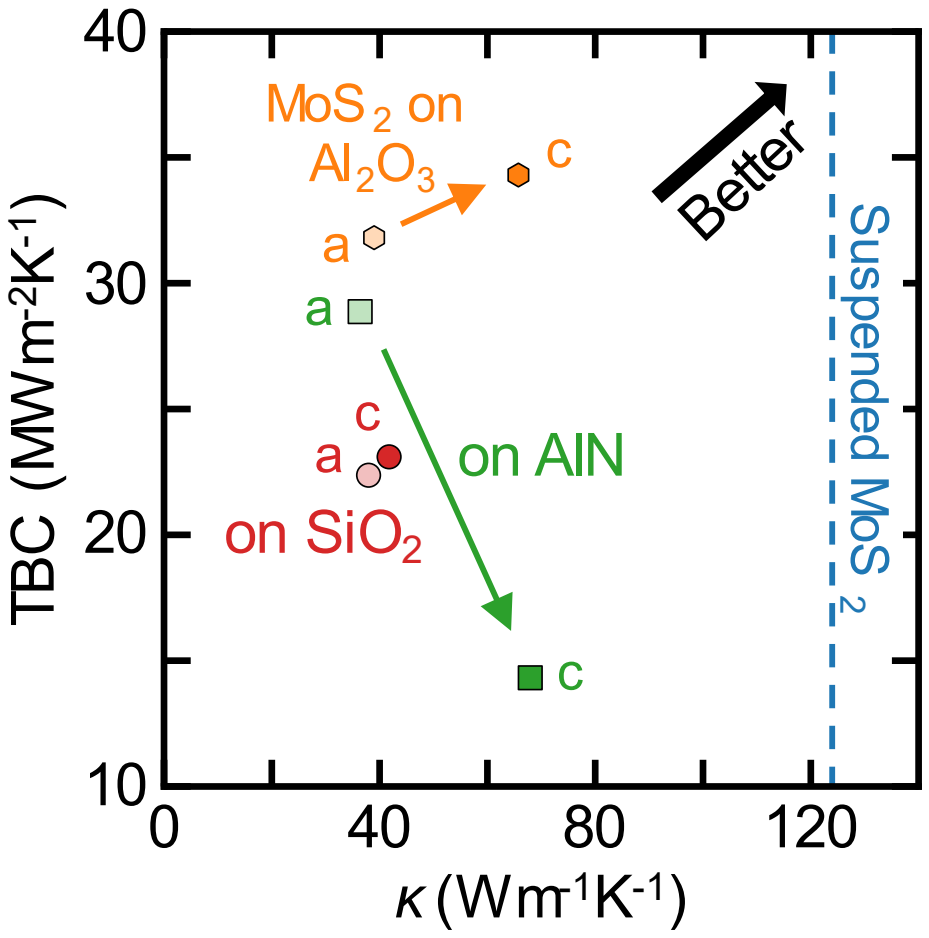

**Figure 5**: A scatter plot showing both the thermal boundary conductance (TBC) and thermal conductivity ($\kappa$) of $MoS_2$ supported by crystalline and amorphous $SiO_2$ (red), AlN (green), and $Al_2O_3$ (orange). The black arrow emphasizes that a larger TBC and thermal conductivity will enable better heat removal from $MoS_2$. The green and orange arrows simply connect related calculations. The dashed vertical blue line denotes the thermal conductivity of suspended $MoS_2$.

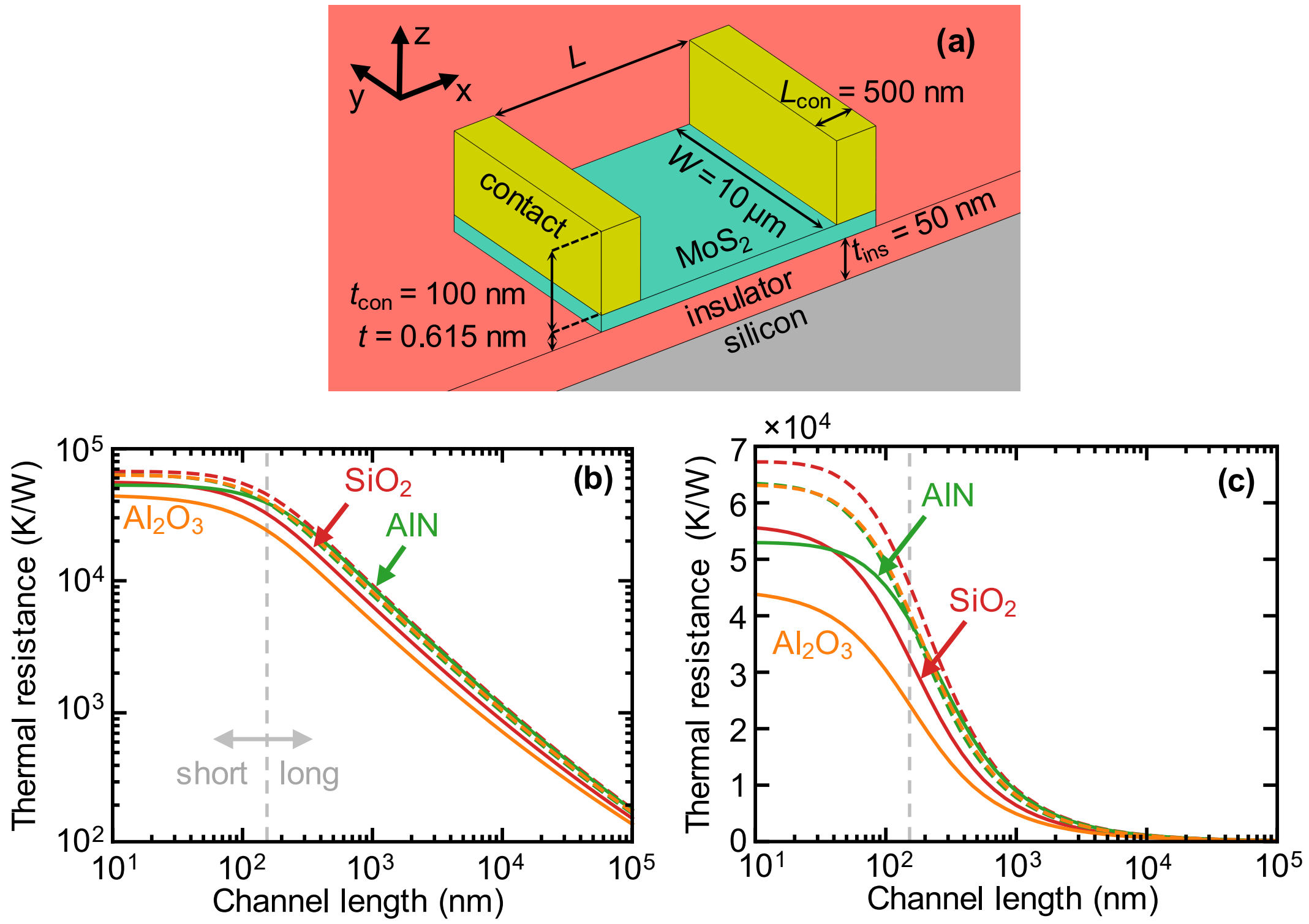

**Figure 6**: **(a)** The back-gated transistor geometry with relevant properties defined for the analytical model (not to scale). Note that this geometry is similar to semiconductor-on-insulator (SOI) transistors, but with the semiconductor at its atomically thin limit with monolayer $MoS_2$. **(b)** The thermal resistance of a back-gated $MoS_2$ transistor, plotted as a function of channel length for different insulator materials, all on a silicon wafer substrate as is often the case in simple proof-of-concept experimental transistors. Solid and dashed lines correspond to the crystalline and amorphous phases of the insulators, respectively, with appropriate $MoS_2$ TCs and TBCs used from our calculations. **(c)** The same data, plotted with a linear-scale vertical axis, to highlight the thermal resistance variation for short-channel transistors. The gray, vertical dashed lines in **(b)** and **(c)** denote the approximate transition between long- and short-channel regimes (*x*-direction).

# Supplementary Information

## Substrate-Dependence of Monolayer $MoS_2$ Thermal Conductivity and Thermal Boundary Conductance

Alexander J. Gabourie[1], Çağıl Köroğlu[1], and Eric Pop[1,2,3,*]

[1]Department of Electrical Engineering, Stanford University, Stanford, CA 94305, USA

[2]Department of Materials Science & Engineering, Stanford University, Stanford, CA 94305, USA

[3]Precourt Institute for Energy, Stanford University, CA 94305, USA

*Contact: epop@stanford.edu

### S1. Lennard-Jones Parameters

We model the van der Waals interactions using the Lennard-Jones (LJ) potential, which is defined as

$$V = 4\epsilon\left[\left(\frac{\sigma}{r}\right)^{12} - \left(\frac{\sigma}{r}\right)^{6}\right], \tag{S1}$$

where $\epsilon$ is the potential well depth, $r$ is the distance between atoms, and $\sigma$ gives $V(r=\sigma) = 0$. For different-atom pairs, the Lorentz-Berthelot mixing rules are used to determine the parameters [i.e., $\epsilon_{AB} = (\epsilon_{AA}\,\epsilon_{BB})^{1/2}$ and $\sigma_{AB} = (\sigma_{AA} + \sigma_{BB})/2$]. The LJ parameters for $MoS_2$ are from the REBO-LJ potential definition [1] and the LJ parameters for other materials are from the Universal Force Field [2]. The single-species LJ parameters are listed in the table below. Application of the mixing rules is left for the reader.

| **Element** | **$\epsilon$** (meV) | **$\sigma$** (Å) |
|---|---|---|
| Mo | 0.58595 | 4.20 |
| S | 20.0 | 3.13 |
| B | 7.80554 | 3.63754 |
| N | 2.99212 | 3.26069 |
| Si | 17.43237 | 3.82641 |
| O | 2.60185 | 3.11815 |
| Al | 21.8989 | 4.00815 |

**Table S**1: The per-element LJ parameters for all species used in this work.

For structures with smooth atomic surfaces (i.e., those with crystalline AlN, $Al_2O_3$, and *h*-BN), the $MoS_2$ can slide on the substrate. Both thermostats and a driving force from the HNEMD method will cause this sliding. For the HNEMD method, the driving forces can accelerate $MoS_2$ laterally, resulting in a divergent thermal conductivity (TC). To fix this problem, we fix the neighbor list during simulation [3]. This effectively tethers the $MoS_2$ to its starting location with an LJ-like bond. Unfortunately, this LJ-like tether induces lateral, sliding oscillations of the $MoS_2$ on the smooth substrates. Based on the heat current expressions [4,5], these oscillations may change the TC; however, we do not find this to be the case. We test an independent set of simulations where we remove the center of mass velocity from each atom and then calculate TC, finding that these lateral oscillations cancel out and do not affect the total TC or the spectral TC of $MoS_2$ on different substrates.

## S2. Structure Creation

We begin by generating an $MoS_2$ sheet and crystalline (c-) $SiO_2$ (quartz), AlN, and $Al_2O_3$ (corundum) substrates such that an interface between them would only result in small stresses. We relax the $MoS_2$ structure for 5 ns using the constant atom number, pressure, and temperature (NPT) ensemble and use the time-averaged lateral dimensions from the last 1 ns to determine the final size of the $MoS_2$ sheet. The relaxed $MoS_2$ area is $10.8 \times 11$ nm$^2$, and we modify the substrates to fit those exact dimensions.

We generate c-$SiO_2$, c-AlN, and c-$Al_2O_3$ using the NanoLab in QuantumATK version P-2019.03 [6]. Each crystalline substrate has a hexagonal lattice structure, which we transform to be orthogonal and repeat to best fit the lateral dimensions of the $MoS_2$ sheet. Residual differences between each substrate and $MoS_2$ are removed by introducing stress in each crystalline substrate. We choose the (0,0,0,1) crystal plane for c-AlN and c-$Al_2O_3$ and the $(1,1,\overline{2},0)$ plane for c-$SiO_2$ to interface with $MoS_2$, which we find to be stable choices. Other surface orientations resulted in either significant rearrangement of atoms at the interface, surface atoms being ejected from the substrate during simulations, or energy drift when running in the NVE ensemble. Interestingly, the thermal conductivity (TC) and spectral TC of $MoS_2$ on both (0,0,0,1) and $(1,1,\overline{2},0)$ crystal planes of c-$SiO_2$ are quantitatively and qualitatively similar (i.e., within error bars), despite the instability of the (0,0,0,1) plane. The c-$SiO_2$, c-AlN, and c-$Al_2O_3$ substrates, respectively, have 23,400 atoms, 33,600 atoms, and 35,800 atoms at thicknesses of 2.7 nm, 2.95 nm, and 2.5 nm.

We create the amorphous (a-) substrates of a-$Al_2O_3$ and a-AlN through an anneal starting from their crystalline counterparts. To create a-$Al_2O_3$, we follow the same procedure as in Ref. [7]. For a-AlN, we find that the long quenching stage of the same procedure allows the material to partially recrystallize. Instead, we rapidly quench the AlN at a rate of $2\times10^{13}$ K s$^{-1}$ with a time step of 0.1 fs until 1000 K then slowly cool from 1000 K to 300 K with a time step of 0.2 fs at $2.8\times10^{12}$ K s$^{-1}$. This adjustment successfully creates a-AlN. We add a vacuum to each structure after the anneal and run a stability check. This check consists of a temperature ramp to 800 K, a hold at 800 K, and a ramp down to 150 K. Each section runs for 250 ps and uses a time step of 0.5 fs. All atoms that are ejected from the surfaces during the stability check are removed from the final amorphous structures. The final a-$Al_2O_3$ structure has 35,878 atoms with a thickness of 2.88 nm and the final a-AlN structure has 33,596 atoms at a thickness of 3.5 nm.

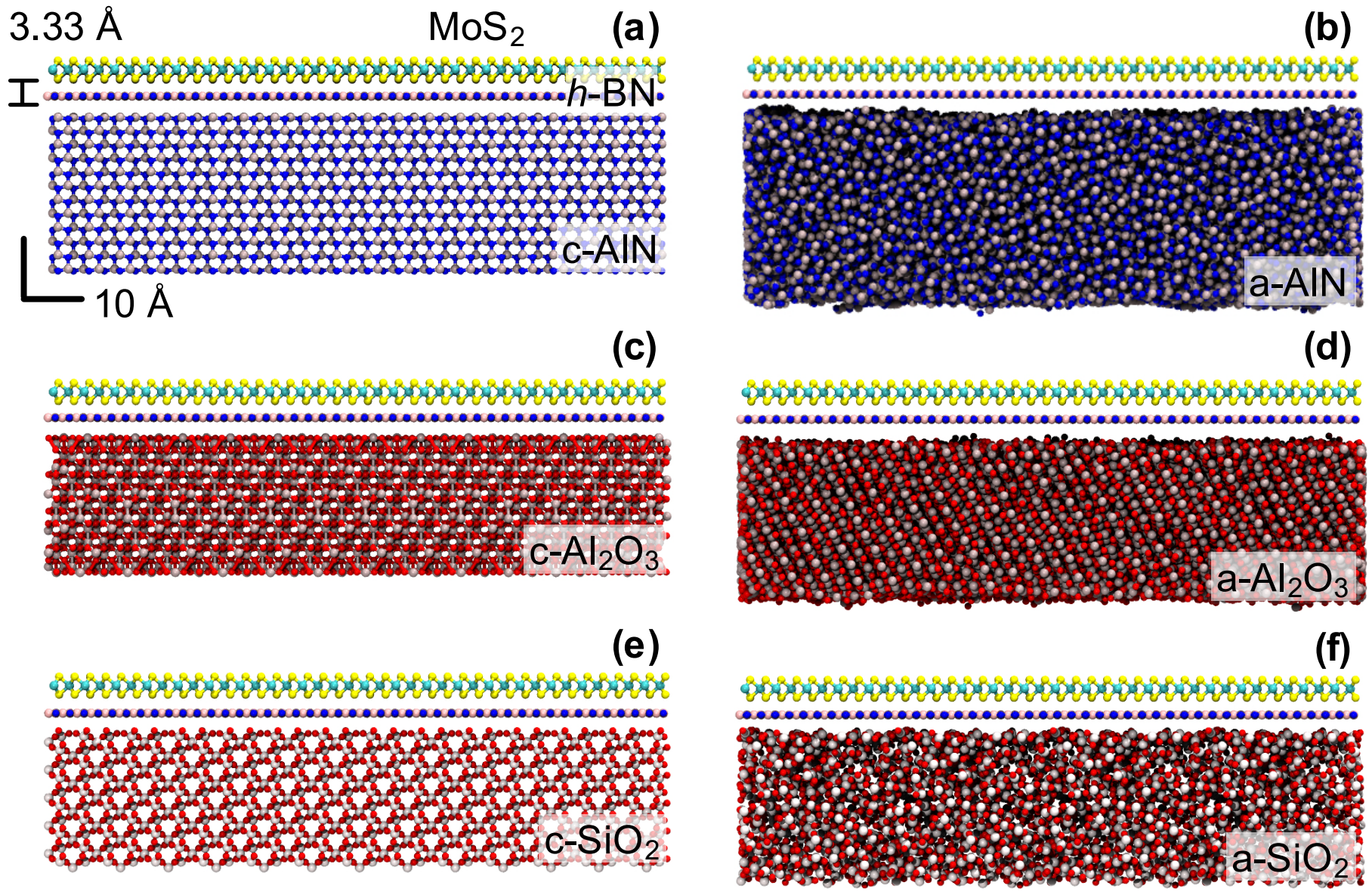

**Figure S1**: Visualizations of the structures used for supported $MoS_2$ thermal conductivity and thermal boundary conductance calculations when there is an *h*-BN interlayer.

We find neither of these procedures, or the popular procedure from Ref. [8], work to rapidly create a viable a-$SiO_2$ structure when starting from our c-$SiO_2$ structure, with diverging energies at any time step when simulating in the constant atom number, volume, and energy (NVE) ensemble. To solve this problem, we use a different c-$SiO_2$ cube, with each lateral dimension one quarter of those for $MoS_2$ (1/16$^{th}$ of the area). We run an ensemble of ten independent anneals (inspired by the notes in the tutorial of Ref. [9]) and test each with an NVE simulation to check for energy drift. We find most structures fail this check. One a-$SiO_2$ realization was viable when using a time step of 0.2 fs. (Use of this time step was critical for nonequilibrium molecular dynamics simulations and thermal boundary conductance calculations.) We repeat the viable a-$SiO_2$ '*cell*' to match the lateral dimensions of $MoS_2$. We also run the same stability check as the other amorphous substrates but with a time step of 0.2 fs. The final a-$SiO_2$ structure has 21,488 atoms is 2.7 nm thick.

Finally, to create supported $MoS_2$ structures, we use the interface builder in QuantumATK [10]. We interface $MoS_2$ with each of the amorphous and crystalline substrates. These structures are run through an energy minimizer to set the proper van der Waals distances between $MoS_2$ and its substrate. The resulting structures can be seen in Figs. 1 (a)-(f) in the main text. We also build structures with an *h*-BN layer between $MoS_2$ and the substrate. For these structures, we complete the interfacing process twice: once for *h*-BN with a substrate and again for $MoS_2$ with *h*-BN on a substrate. Structures with *h*-BN interlayers are in Fig. S1.

**S3. Homogenous Nonequilibrium Simulations**

Calculations of the TC with the homogeneous nonequilibrium molecular dynamics simulations (HNEMD) and the frequency-dependent (spectral) TC with the spectral heat current method (SHC) are concurrent [5,11,12]. The simulation protocol consists of an equilibration and production step, both using a time step of 0.5 fs. The equilibration step runs each structure for 500 ps in the constant atom number, volume, and temperature (NVT) ensemble using the Nosè-Hoover chain thermostat [13]. Typically, equilibration is run in the NPT ensemble to allow structures to change volume and relax to a configuration natural for the potentials in use [14]; however, since the substrates in all supported $MoS_2$ simulations have many more atoms than $MoS_2$, they would dictate the final size of the simulation cell during an NPT run and $MoS_2$ would be strained differently for different substrates. Since the TC of $MoS_2$ has been shown to be sensitive to strain [15,16], we fix the lateral dimensions of $MoS_2$ for consistency across simulations.

The production step depends on the system being considered with two parameters being varied: production time and driving force parameter $F_e$ (which applies the driving force in only the *x*-direction in this study). For suspended $MoS_2$, a consequence of the NVT equilibration step is that the TC from the HNEMD method is more difficult to converge. While previous studies have found that a production time of $t_s \leq 10$ ns and a driving force parameter of $F_e = 0.2$ μm$^{-1}$ [7,17] is sufficient, we find that suspended $MoS_2$ requires $t_s = 25$ ns and $F_e = 0.08$ μm$^{-1}$ to converge. For all other structures (i.e., supported $MoS_2$), $t_s = 5$ ns and $F_e = 0.12$ μm$^{-1}$ work well. For concurrent SHC calculations, we use a sampling period of 2 fs as well as a maximum correlation time of 500 fs for all structures. The results of the HNEMD and SHC calculations are from 100 independent simulations of each structure, except those with *h*-BN, which are from 50 independent simulations. Note that the integral over the spectral TC curves recovers the TC from the HNEMD method. The results of the HNEMD TC calculations are shown Fig. S2 and Fig. S3 and the full set of results for spectral TC with the SHC method are shown in Fig. S4 and Fig. S5. Additional spectral TC comparisons can be found in Fig. S6 and Fig. S7.

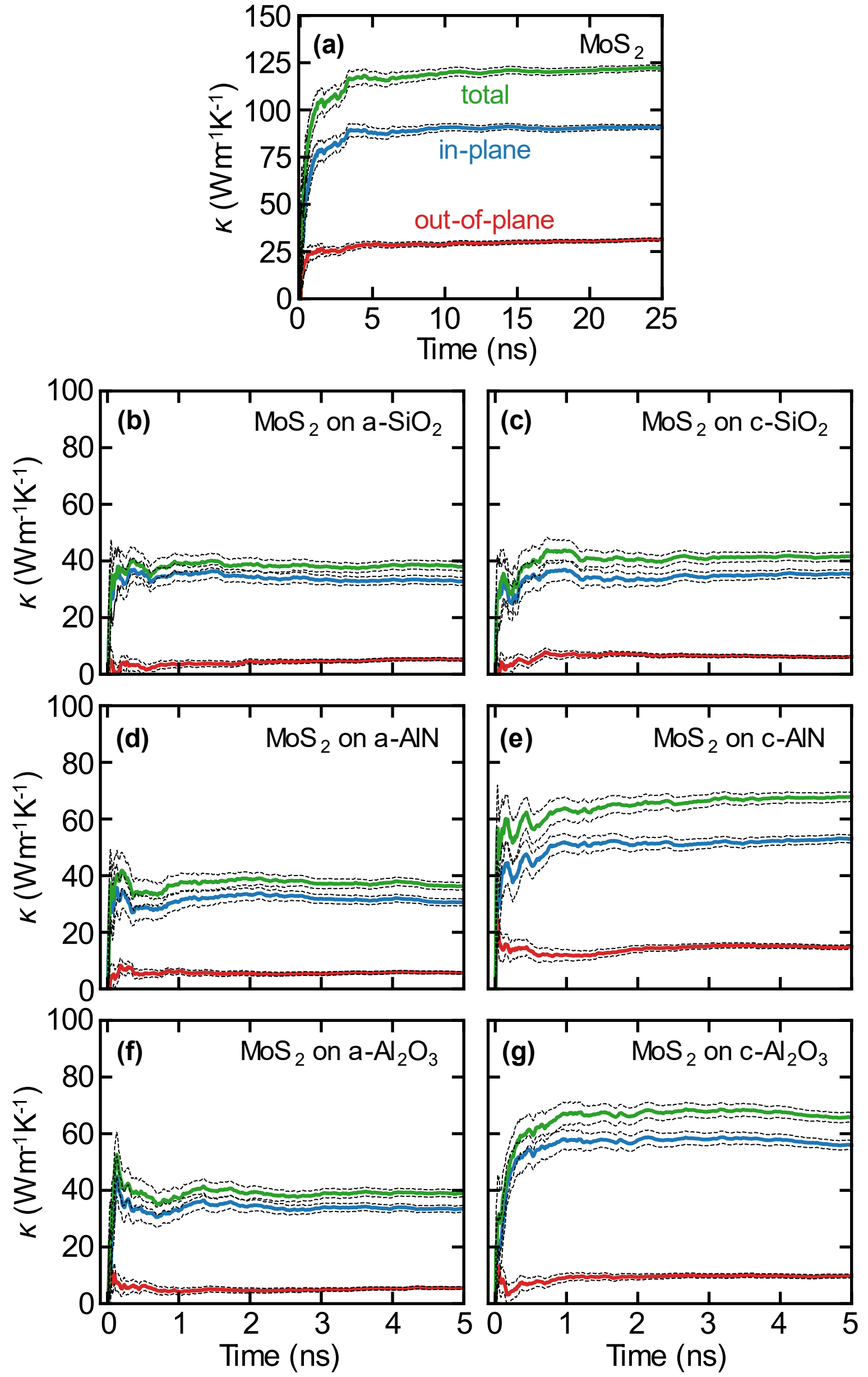

**Figure S2**: The running thermal conductivity of suspended $MoS_2$ as well as $MoS_2$ supported by amorphous and crystalline $SiO_2$, AlN, and $Al_2O_3$ using the HNEMD method. Contributions from in-plane (blue lines) and out-of-plane (red lines) atomic motion as well as the total thermal conductivity (green lines) are shown in each panel. The dashed black lines denote the standard error.

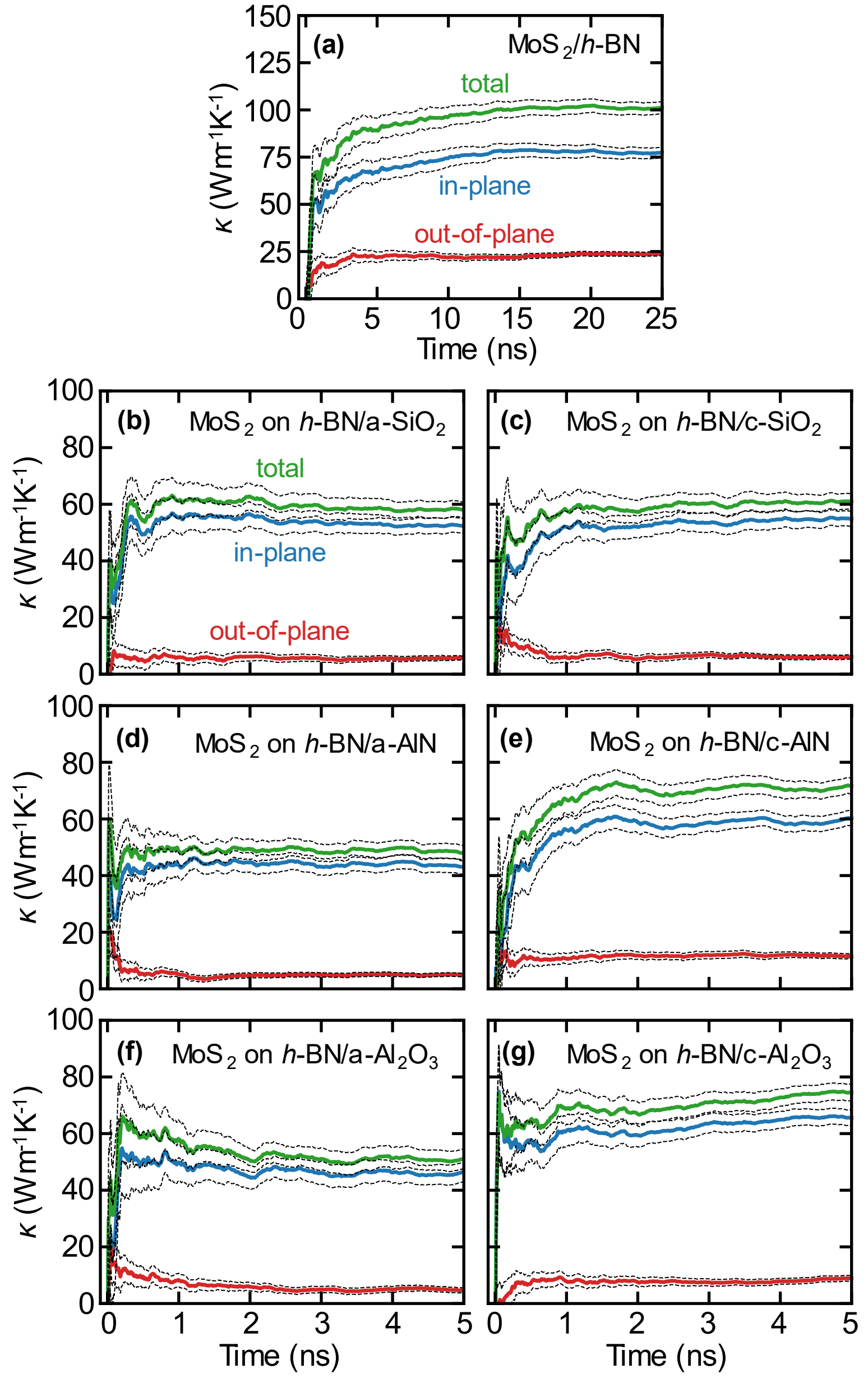

**Figure S3**: **(a)** The running HNEMD thermal conductivity of $MoS_2$ in a heterostructure with a single-layer of *h*-BN. **(b)-(g)** The running HNEMD thermal conductivity $MoS_2$ supported by amorphous and crystalline $SiO_2$, AlN, and $Al_2O_3$ with a single interlayer of *h*-BN. Contributions from in-plane (blue lines) and out-of-plane (red lines) atomic motion as well as the total thermal conductivity (green lines) are shown in each panel. The dashed black lines denote the standard error.

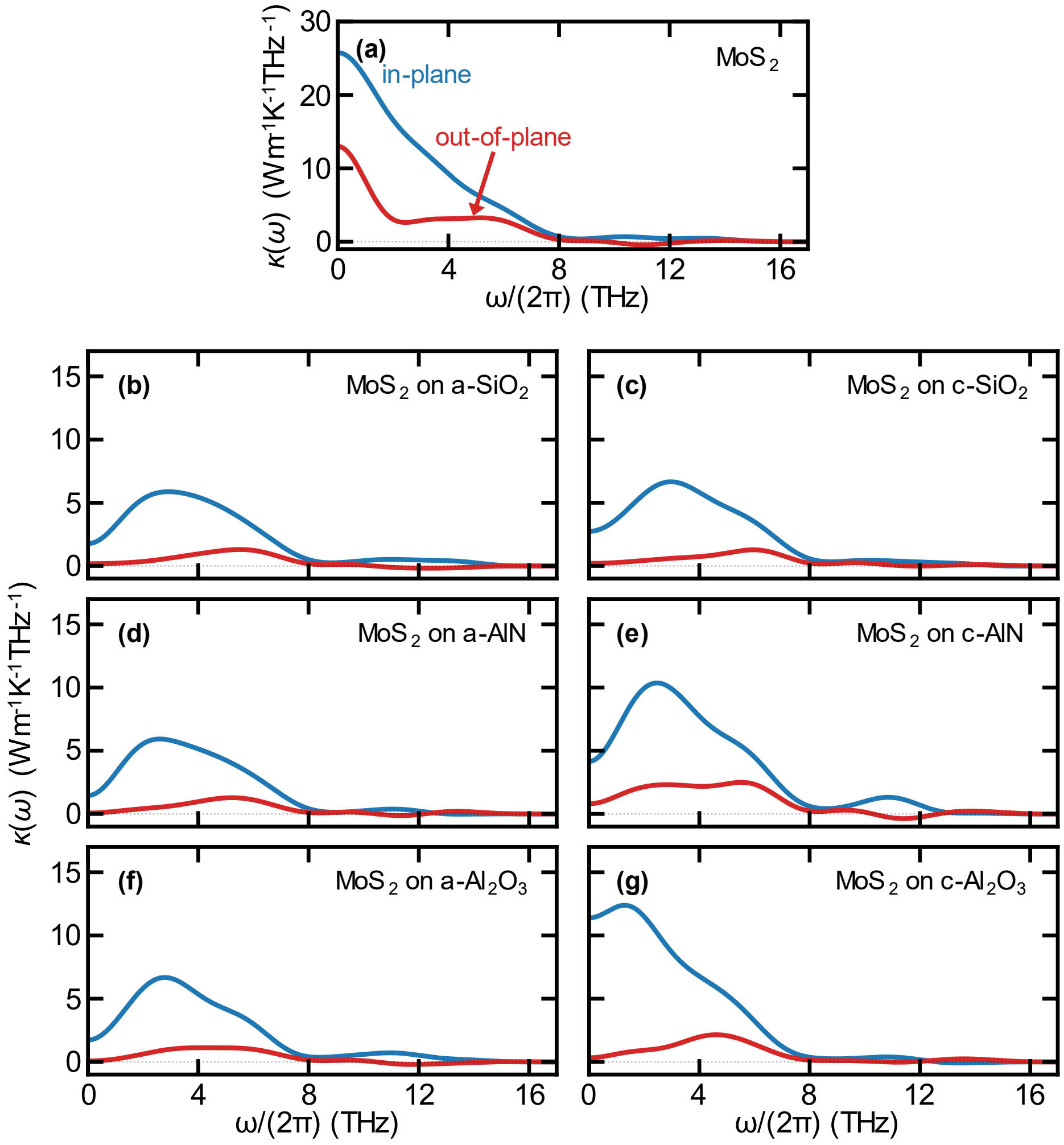

**Figure S4**: The spectral thermal conductivity of suspended **(a)** and supported **(b)-(g)** $MoS_2$ calculated with the SHC method over all $MoS_2$ phonon frequencies. The blue lines denote contributions from in-plane atomic motion and the red lines denote contributions from out-of-plane atomic motion. Note the different *y*-axis scales between **(a)** and the other subplots.

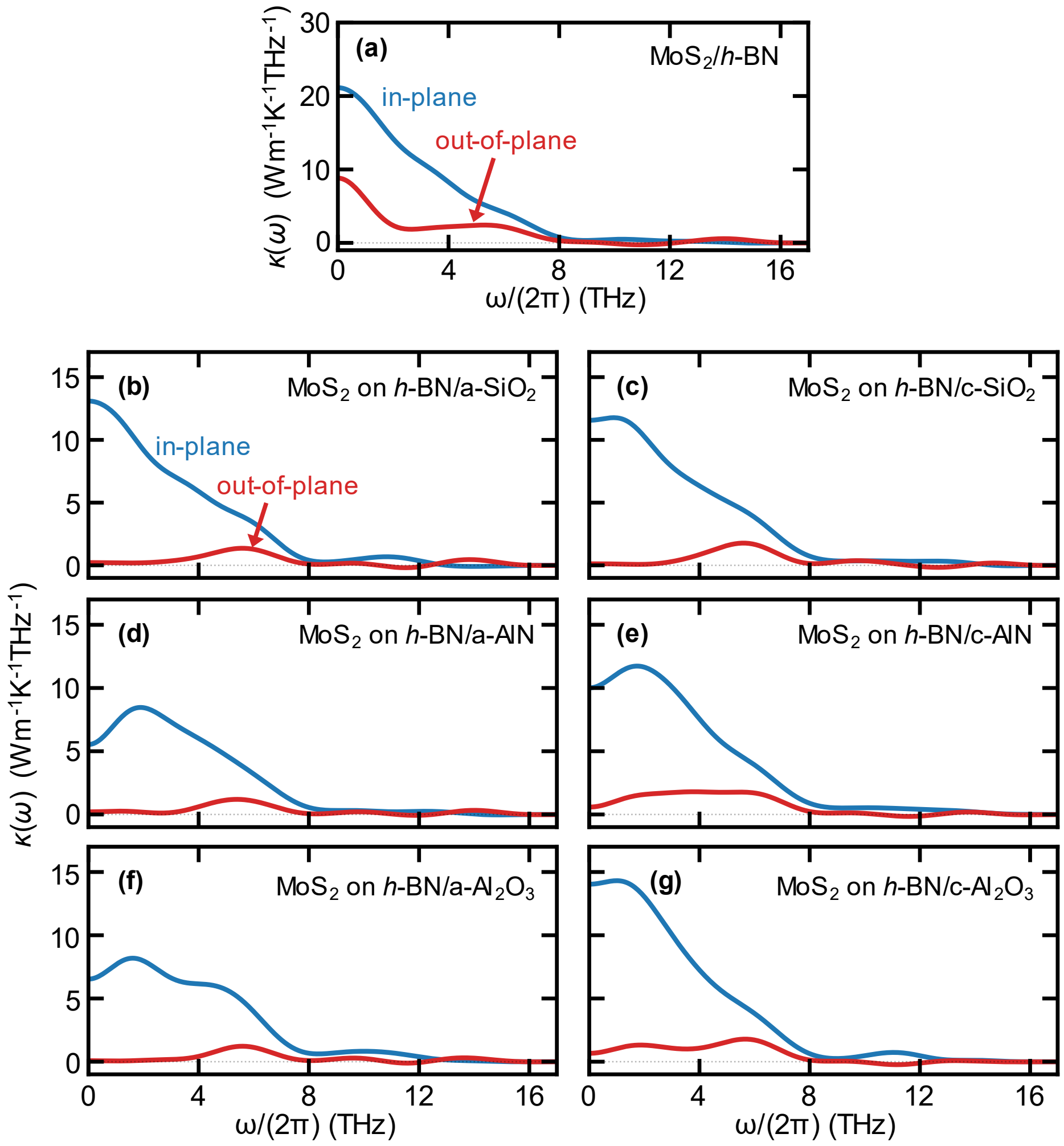

**Figure S5**: The spectral thermal conductivity of $MoS_2$ over all phonon frequencies using the SHC method for an $MoS_2$/*h*-BN heterostructure **(a)** and $MoS_2$ supported by substrates that are capped by a single layer of *h*-BN on top **(b)-(g)**. The contributions from in-plane and out-of-plane atomic motion are shown in the blue and red lines, respectively. Note the different *y*-axis scales between **(a)** and the other subplots.

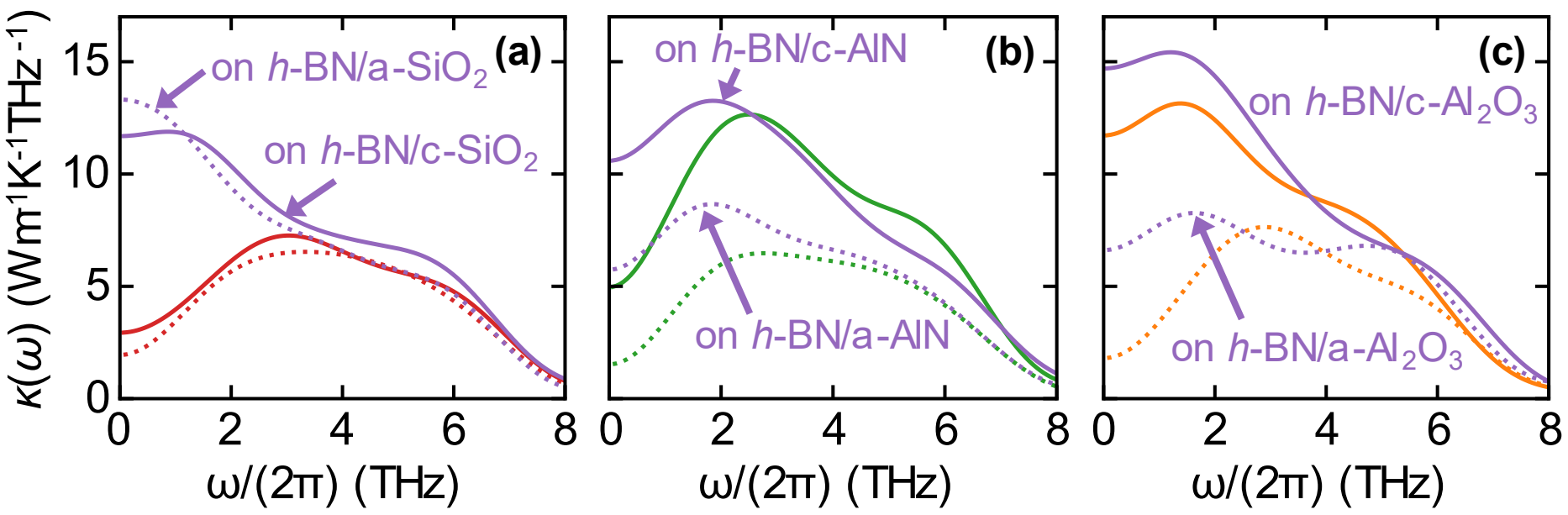

**Figure S6**: A comparison of the total spectral thermal conductivity $\kappa(\omega)$ of $MoS_2$ over its acoustic modes on each of the substrate materials with and without an *h*-BN interlayer. The dotted and solid lines denote amorphous and crystalline substrates, respectively. Results without *h*-BN are shown in red for $SiO_2$ in **(a)**, green for AlN in **(b)**, and orange for $Al_2O_3$ in **(c)**. The purple curves are for structures with *h*-BN interlayers. Note how $\kappa(\omega)$ contributions improve at low frequencies (i.e., < 3 THz) when *h*-BN is added.

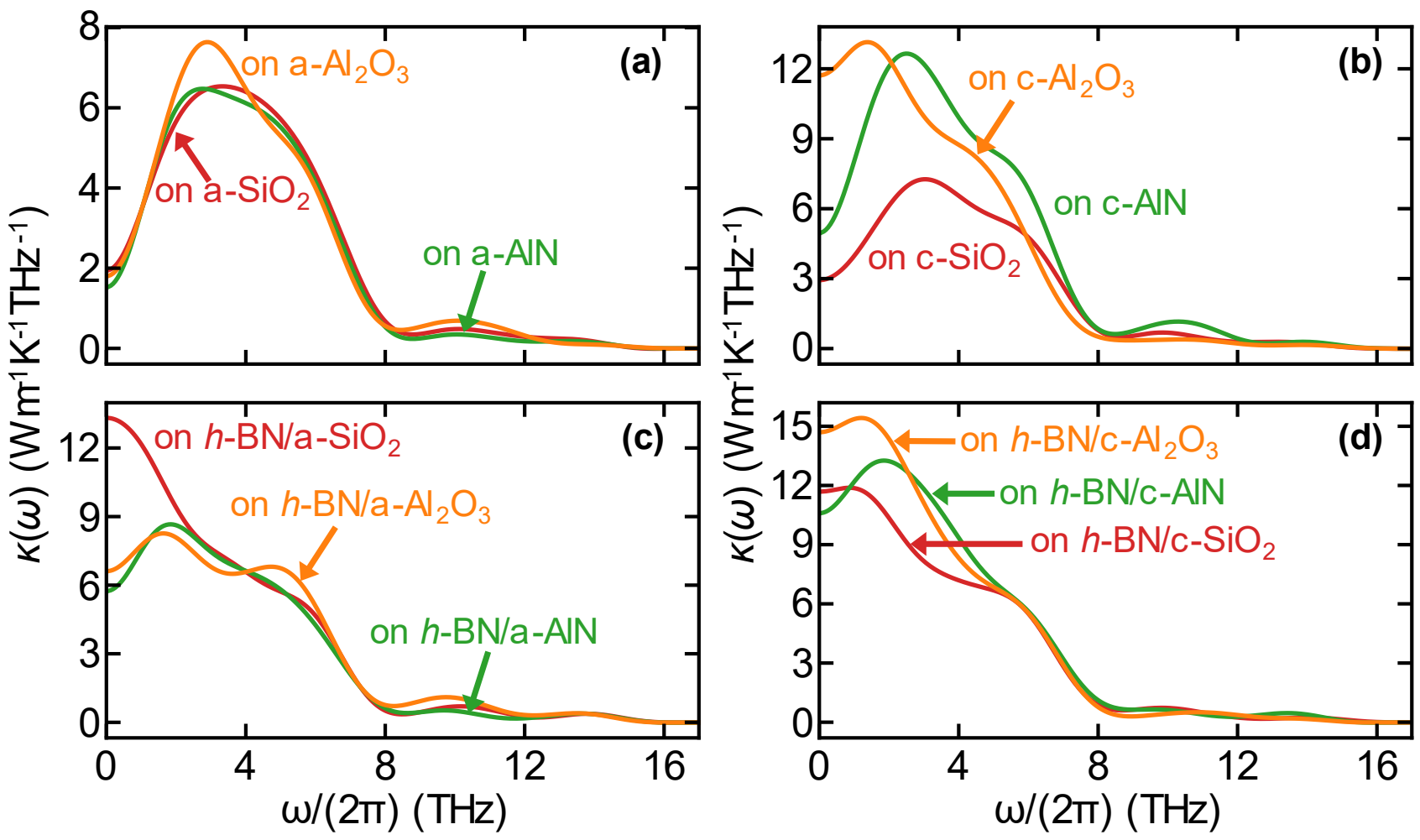

**Figure S7**: The total spectral thermal conductivity of $MoS_2$ when supported by **(a)** amorphous substrates, **(b)** crystalline substrates, **(c)** amorphous substrates with an *h*-BN layer on top, and **(d)** crystalline substrates with an *h*-BN layer on top. Note the different *y*-axis scales in each subplot.

## S4. Vibrational Density of States Calculations

The vibrational density of states (VDOS) represents the density of normal modes (or phonons in crystals) at a particular time-varying frequency. We calculate the VDOS from the Fourier transform (specifically the discrete cosine transform) of the velocity autocorrelation function [7,18,19]. From a thermal perspective at interfaces, it can detail what types of interactions may occur between materials and has been used to analyze supported TC and thermal boundary conductance (TBC) [7,20]. While the VDOS is lacking spatial information and therefore does not have information on the momentum of normal modes at each frequency, some spatial information can be inferred from it as more normal modes imply a greater spread of crystal momenta. In Fig. S8 and Fig. S9 below, we show additional comparisons of the VDOS not shown in the main text.

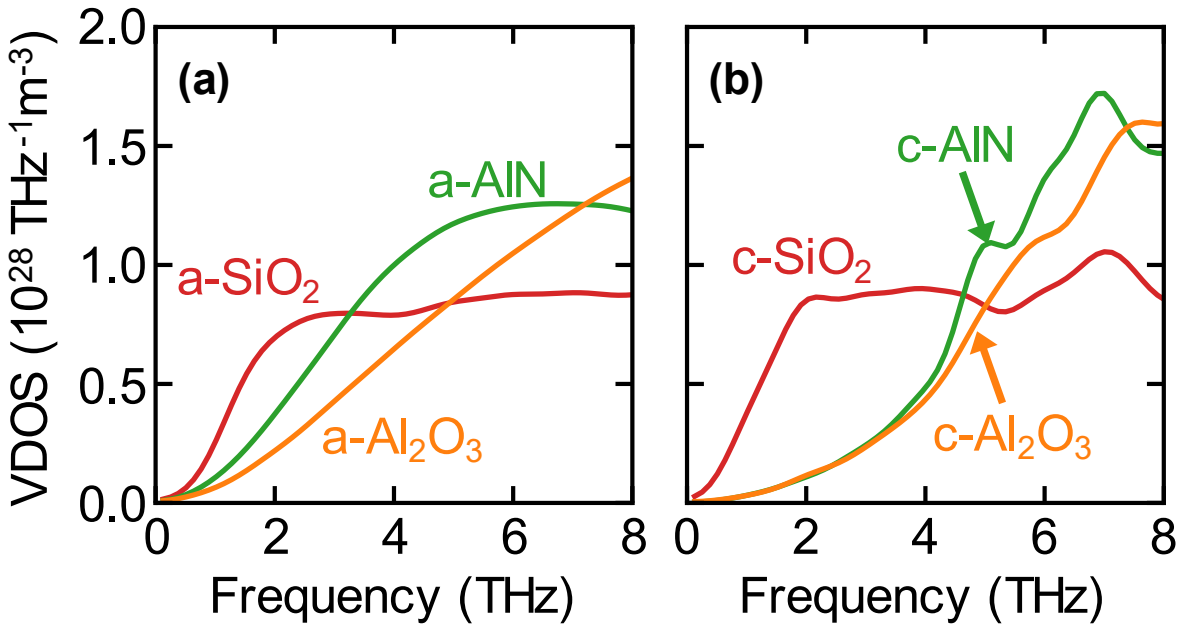

**Figure S8**: A comparison of the VDOS of **(a)** amorphous substrates and **(b)** crystalline substrates.

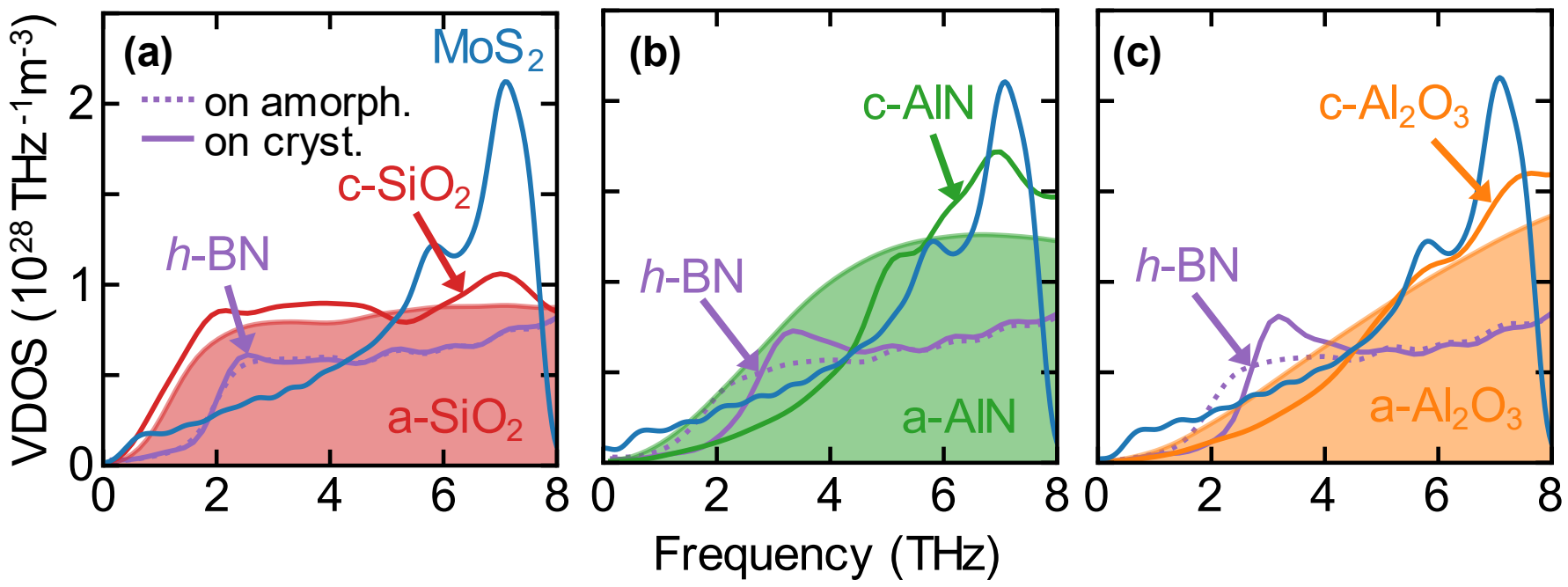

**Figure S9**: The VDOS of $MoS_2$ on *h*-BN-capped amorphous and crystalline substrates. The dashed and solid purple lines denote *h*-BN VDOS when on the corresponding amorphous and crystalline substrates, respectively. All VDOS curves for $SiO_2$ are in **(a)**, all for AlN are in **(b)**, and all for $Al_2O_3$ are in **(c)**.

### S5. van der Waals Force Spectrum

Previous theoretical studies have used interfacial pressure or force to analyze the thermal boundary conductance (TBC) between materials, including for systems with 2D/3D interfaces [21-23]. Here, we also leverage this interfacial force to better understand the TC and TBC of supported $MoS_2$ by isolating the vdW force each substrate applies to $MoS_2$ via the LJ potential. Any type of interaction between a substrate and $MoS_2$ must be facilitated through LJ interactions. The simulation protocol for data collection is as follows: First, we equilibrate each system in the NVT ensemble for 500 ps. Next, we run an additional 2 ns in the NVT ensemble, sampling the cumulative LJ force on $MoS_2$ every 10 fs. To compare to the spectral TC figures, we project the vdW force data into the frequency domain by using the expression

$$f_z^{\mathrm{LJ}}(\omega_k) = \left| \sum_{i,n} F_{z,i}^{\mathrm{LJ}}[n] e^{-j\omega_k \frac{n}{M}} \right|, \tag{S2}$$

where $f_z^{\mathrm{LJ}}$ is the magnitude of the $z$-component force $MoS_2$ feels from the substrate at frequency $\omega_k$ and $F_{z,i}^{\mathrm{LJ}}[n]$ is the $z$-component force the $i$th $MoS_2$ atom feels from the substrate at time $n$. The summation is over all $N$ atoms and $M$ frequencies (i.e., $k \in [0, M\text{-}1]$). The result of these calculations for $MoS_2$ in every structure, can be seen in Fig. S10 and Fig. S11 below. The $x$- and $y$-component LJ forces are small compared to the $z$-component and are not shown here.

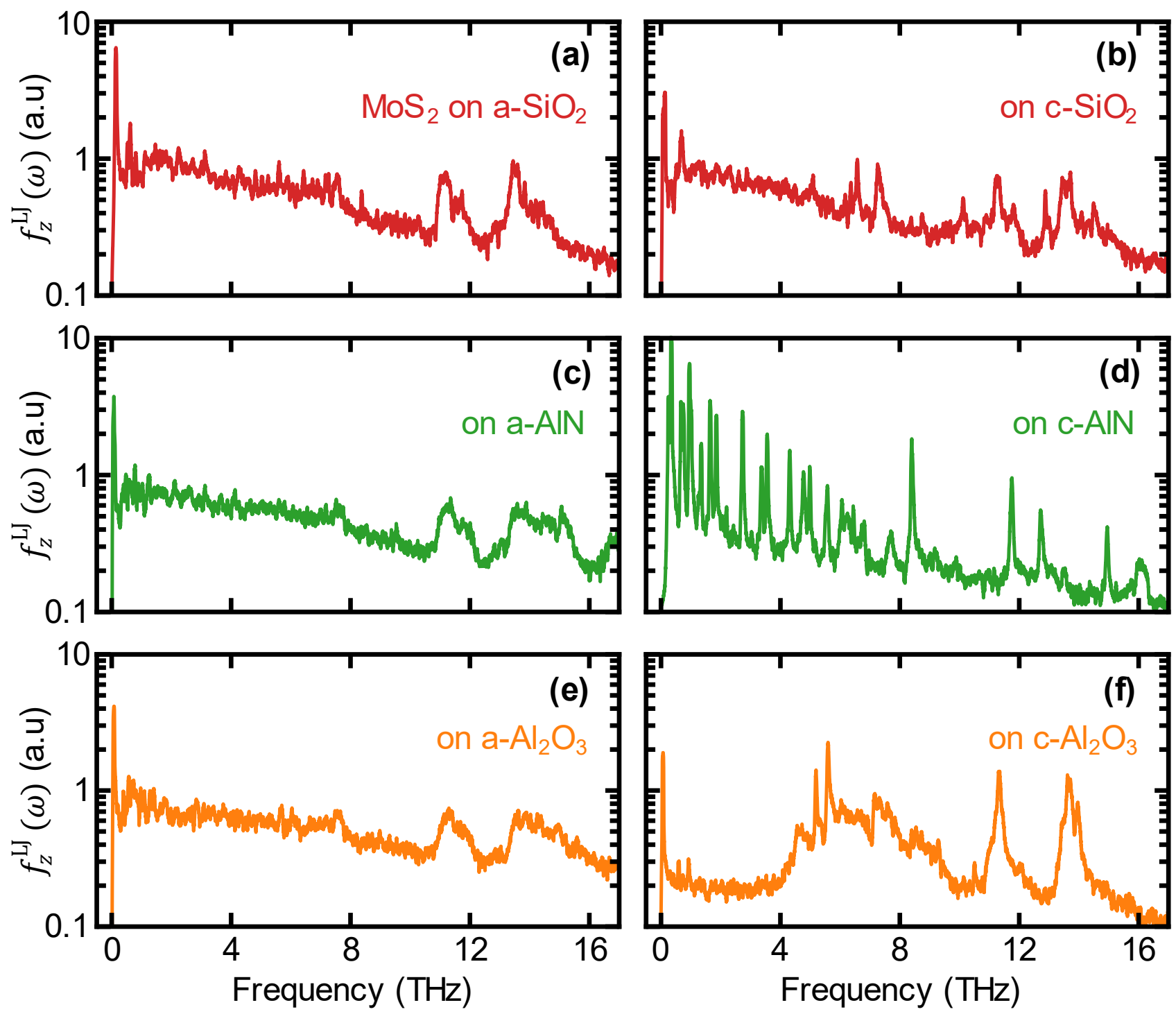

**Figure S10**: The force spectrum on $MoS_2$ through vdW forces (via the LJ potential) from each substrate.

We see similar vdW forces [Figs. S10(a,c,e)] and cumulative vdW forces [Fig. S12(a)] on $MoS_2$ from each of the amorphous substrates. These results support Fig. 2 of the main text where we find the spectral TCs for $MoS_2$ supported by amorphous substrates to also be similar. Interestingly, the forces c-$SiO_2$ [Fig. S10(b)] exert on $MoS_2$ are like those of a-$SiO_2$ [Fig. S10(a)]. This is also in agreement with the comparable VDOS and spectral TC curves we found each to have.

The forces on $MoS_2$ from c-AlN and c-$Al_2O_3$ in Figs. S10(d,f) look very different than from other substrates. When $MoS_2$ is on c-AlN, the vdW force spectrum is comparatively large up to ~3 THz. Its relative strength can be more clearly seen in Fig. S12(b) as the cumulative vdW forces from c-AlN rise quickly in this frequency range. The frequencies of these strong vdW forces matches well with the region where we see severe degradation in the spectral TC contributions of $MoS_2$ on c-AlN in Fig. 2(c) of the main text [or Fig. S7(b)]. (Note that the very sharp vdW force peaks are a due to size effects: the thicker the substrate, the more peaks will be present.)

When $MoS_2$ is on c-$Al_2O_3$, the vdW forces between 0 THz and 4 THz are very small compared to the other substrates. In this frequency range, we see that the spectral TC of $MoS_2$ on c-$Al_2O_3$ is also significantly higher than when on other substrates, closer to that of suspended $MoS_2$. These results, as well as those for $MoS_2$ supported by c-AlN, suggest that the strength of the vdW force is directly related to the degradation of the spectral TC at a given frequency. Between 4 THz and 8 THz, the vdW forces on $MoS_2$ from c-$Al_2O_3$ increase to become comparable or stronger than those from the other substrates. While these forces degrade the TC of $MoS_2$ in this frequency range, their effect on the total TC of $MoS_2$ is reduced as a large majority of the TC contributions for suspended $MoS_2$ are at frequencies below 4 THz. However, the forces in this frequency range may be responsible for the high TBC between $MoS_2$ and c-$Al_2O_3$ as we hypothesize in the main text.

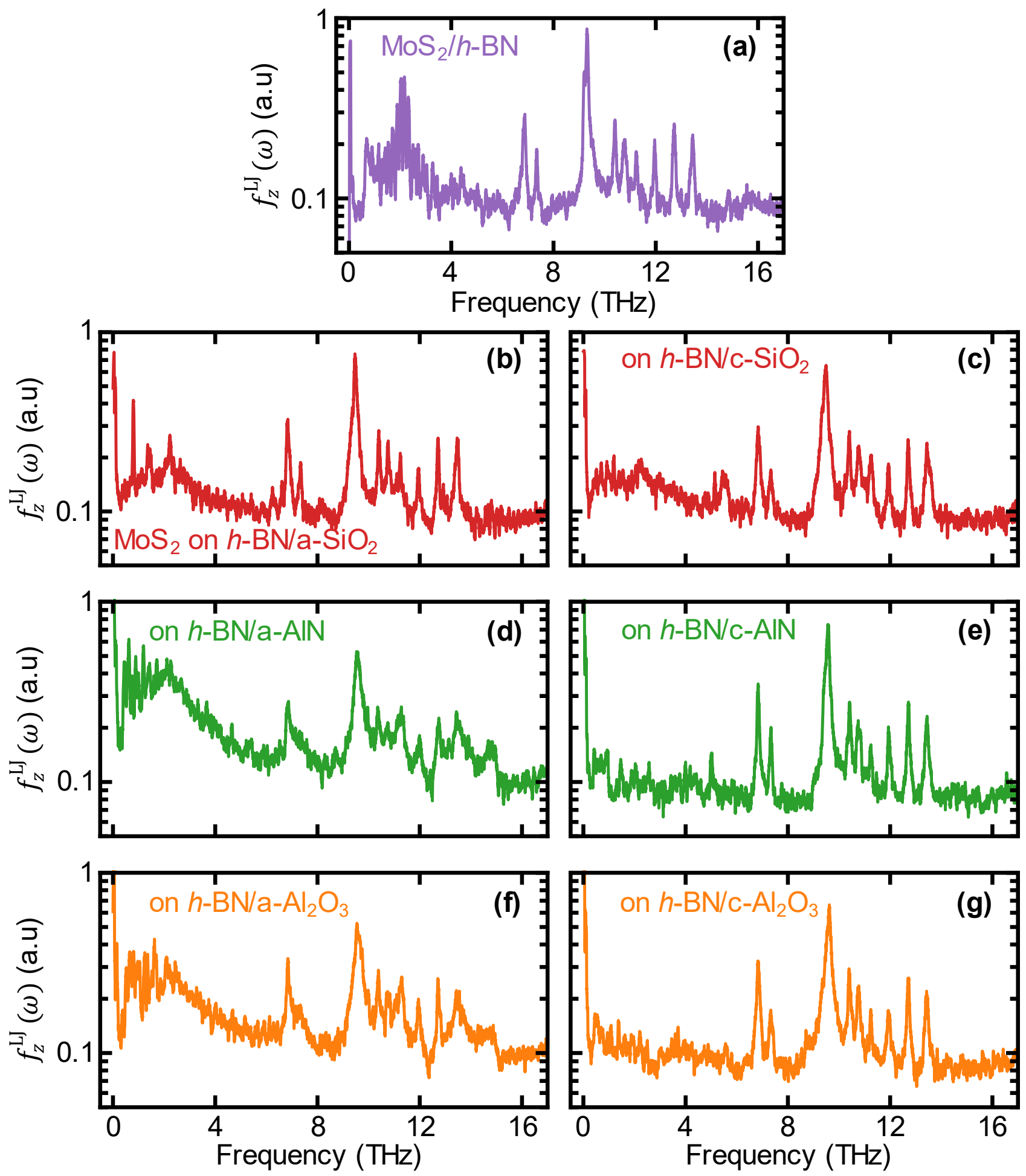

**Figure S11**: **(a)** The force spectrum on $MoS_2$ from *h*-BN in an $MoS_2$/*h*-BN heterostructure. **(b)-(g)** The force spectrum on supported $MoS_2$ when supported there is an *h*-BN interlayer.

For *h*-BN-capped substrates, we see the vdW force spectra on $MoS_2$, shown in Fig. S11, to be much smaller than their non-*h*-BN counterparts (Fig. S10) across most of the spectrum, but especially between 0 THz and 4 THz. The reduction in vdW forces between 0 THz and 4 THz corresponds to a higher spectral TC for $MoS_2$ compared to structures without the *h*-BN interlayer. These spectral TC comparisons can be seen in Fig. S6. The *h*-BN interlayer acts as a barrier, blocking $MoS_2$ from substrate forces and allowing the spectral TC of supported $MoS_2$ to behave closer to the spectral TC seen in Fig. 2(a) for suspended $MoS_2$. This 'blocking' behavior is evidenced by the vdW force spectrum shown in Fig. S11(a), when $MoS_2$ is in an $MoS_2$/*h*-BN heterostructure, as the rest of the vdW force spectra in Fig. S11 look similar to it. Note that the vdW forces are also smaller in the 4 THz to 8 THz range when *h*-BN is present. This result is more important for the c-$Al_2O_3$ substrate as, now that the vdW force 'bump' in this frequency range, seen in Fig. S10(f), is gone, the TBC of $MoS_2$ on *h*-BN/c-$Al_2O_3$ is now similar to the TBC of $MoS_2$ on *h*-BN/c-AlN. This supports the hypothesis that vdW forces in this frequency range are responsible for the large TBC of $MoS_2$ on c-$Al_2O_3$.

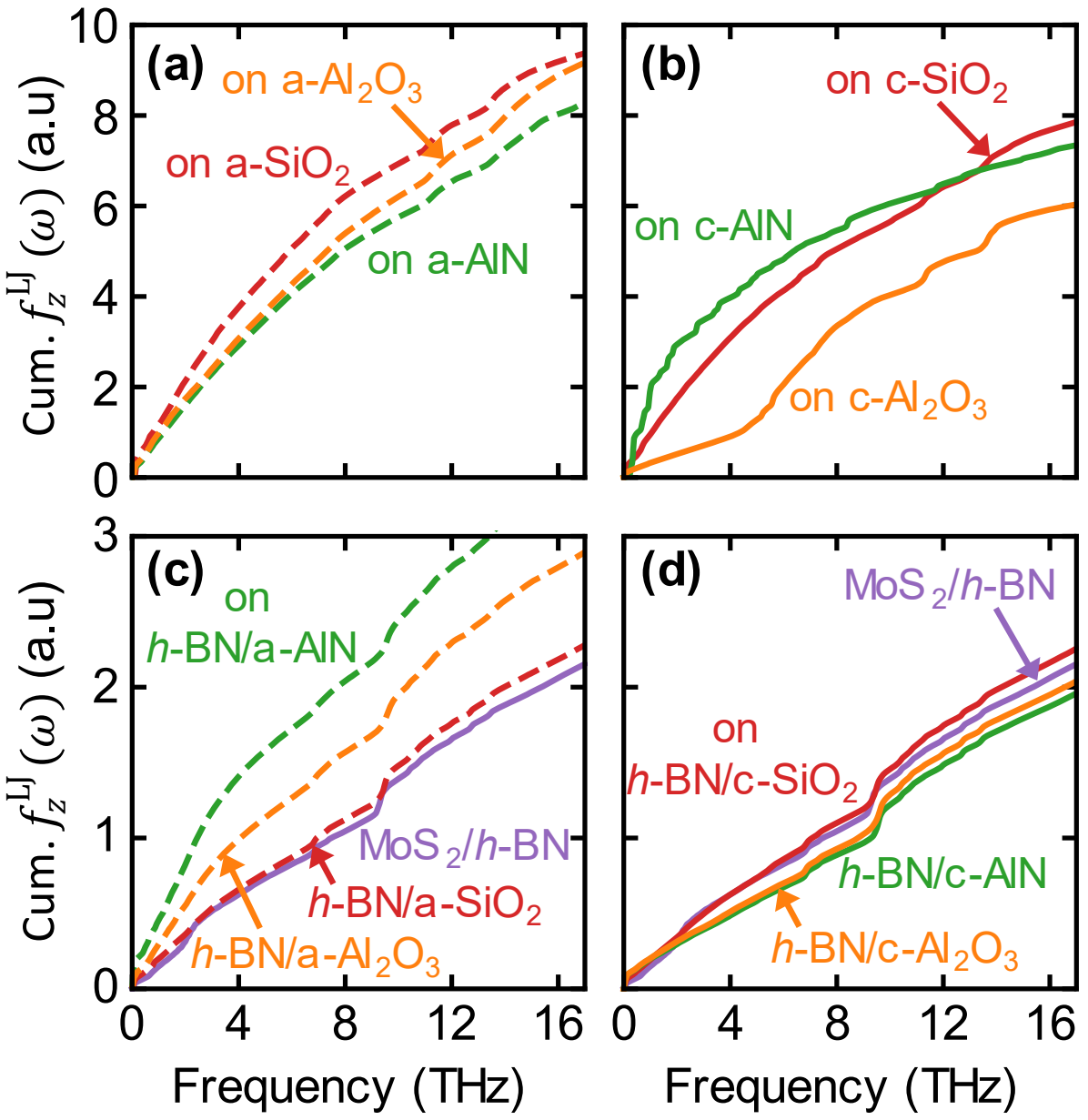

**Figure S12**: The cumulative vdW force spectrum for $MoS_2$ on **(a)** amorphous and **(b)** crystalline substrates as well as on **(c)** amorphous and **(d)** crystalline substrates with an *h*-BN interlayer.

## S6. Nonequilibrium Simulations

To calculate the frequency-dependent (spectral) phonon mean free path (MFP) and the length-dependent TC, we first calculate the thermal conductance $G(\omega)$ using nonequilibrium molecular dynamics (NEMD) simulations. The thermal conductance can be calculated with [12,24]

$$G(\omega) = \frac{2\tilde{K}(\omega)}{\Delta T V}, \tag{S3}$$

where $\tilde{K}(\omega)$ is the Fourier transform of the virial-velocity correlation as defined in Refs. [5,11], $V$ is the volume over which the SHC method is calculated, and $\Delta T$ is the temperature difference between the two temperature reservoirs [25,26].

Before an NEMD simulation, we partition the structure into different regions. We fix a thin layer of 120 atoms in the first ~1.6 Å of the 10.8 nm $x$-dimension and create two ~19.4 Å, 720 atom regions, one at each end of the $x$-direction of the $MoS_2$ sheet, to act as the hot and cold thermal reservoirs. For the remaining parts, we create seven, ~9.7 Å, 360 atom regions for temperature recording. As with other simulations, periodic boundary conditions are used in the lateral directions. A representation of the $MoS_2$ structure and simulation setup can be seen in Fig. S13(a).

The NEMD simulation protocol is as follows: First, each system is equilibrated for 500 ps in the NVT ensemble. Next, we switch to the NVE ensemble and apply local Langevin thermostats [27,28] to the thermal baths [24] for 1 ns to establish a steady-state temperature difference of $\Delta T = 10$ K between them. Holding these conditions, we run for an additional 5 ns, recording the temperatures and the heat transfer in the thermal baths. During this stage, we also calculate the SHC in the center layer of the $MoS_2$. The sample period of the SHC method is 2 fs with a maximum correlation time of 750 fs. Note that the time step was 0.5 fs for all structures except those including a-$SiO_2$ which used a time step of 0.2 fs, and we simulate 3 independent runs for each structure, averaging the results. We show the NEMD results of $MoS_2$ supported by c-$Al_2O_3$ as a representative example for results of temperature profiles, heat transfer in the thermal baths, the time-domain virial-velocity correlation from the SHC method [5], and the thermal conductance $G(\omega)$ [12,24] in Figs. S13(b)-(f).

Finally, we combine the thermal conductance with the spectral TC to calculate the phonon MFP with $\lambda(\omega) = \kappa(\omega)/G(\omega)$ [11] and the length-dependent TC $\kappa(L)$ using Eq. (3) from the main text. Figure 4 in the main text shows the results for $MoS_2$ on substrates without *h*-BN and discusses the result from all structures. Figure S14 shows similar results but for structures with an *h*-BN layer between $MoS_2$ and each substrate.

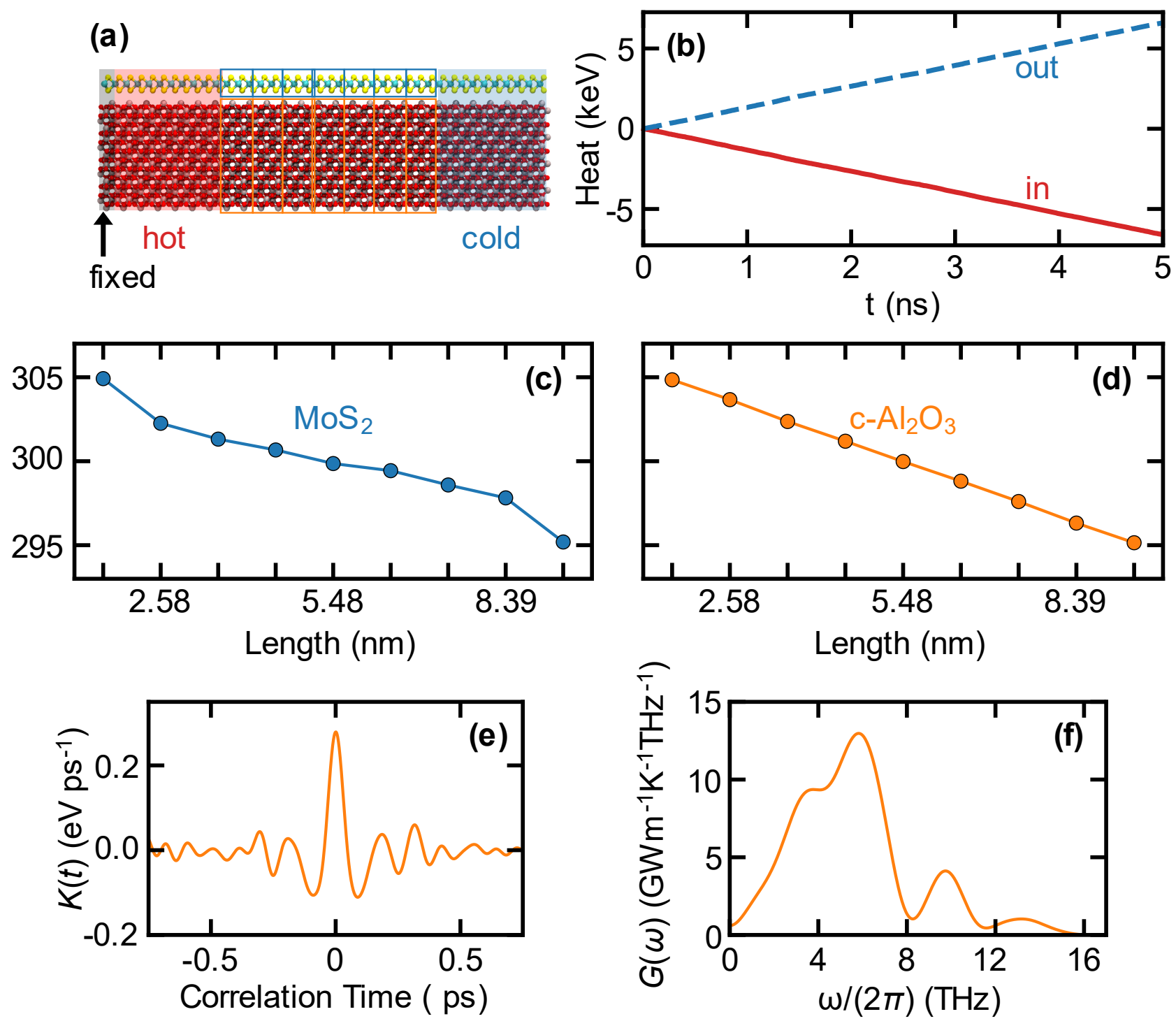

**Figure S13**: **(a)** The NEMD simulation setup for $MoS_2$ supported by c-$Al_2O_3$. The gray region shows the fixed atoms, and the red and blue regions show the hot and cold reservoirs. The blue squares show the sections where the temperature of $MoS_2$ was sampled, and the orange rectangle shows where the temperature of c-$Al_2O_3$ was sampled. **(b)** The total heat coming in or out of the system through the reservoirs. The temperature profiles for $MoS_2$ and c-$Al_2O_3$, during steady-state, are shown in **(c)** and **(d)**, respectively. **(e)** The time-domain virial-velocity correlation (defined in Refs. [5,11]) for $MoS_2$ on c-$Al_2O_3$ **(f)** The spectral thermal conductance of $MoS_2$ on c-$Al_2O_3$ [defined in Eq. (S3)].

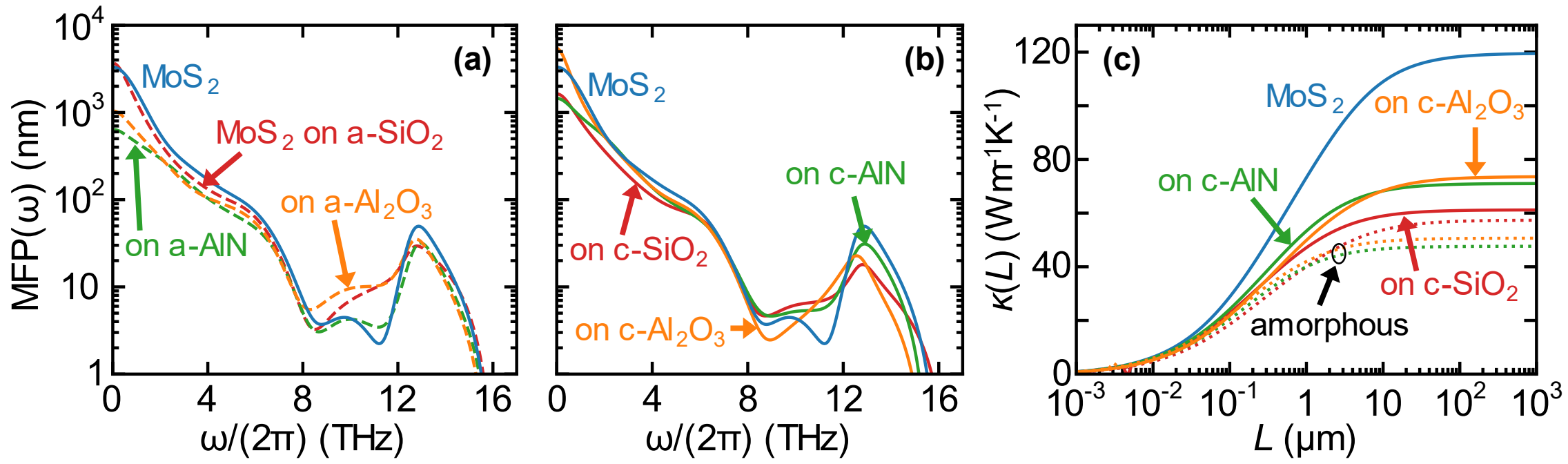

**Figure S14**: The phonon mean free path (MFP) of suspended $MoS_2$ as well as $MoS_2$ supported by *h*-BN-capped **(a)** amorphous and **(b)** crystalline $SiO_2$, AlN, and $Al_2O_3$. **(c)** The length-dependent thermal conductivity of suspended $MoS_2$ and $MoS_2$ supported by all *h*-BN-capped substrates.

### S7. Approach to Equilibrium Method

To calculate the TBC between $MoS_2$ and each substrate, we use the approach to equilibrium MD (AEMD) method [8,29,30]. We begin the simulation protocol with a 100 ps equilibration of the system in the NVT ensemble at 350 K. Next, we switch to the NVE ensemble and apply two Langevin thermostats [27,28]: one for $MoS_2$ set to 400 K and the other for the entire substrate set to 300 K. This step runs for 500 ps to create a stable temperature difference of $\Delta T_0$ = 100 K. Finally, we remove the thermostats and let $MoS_2$ cool into the substrate while tracking the temperature difference $\Delta T(t)$. A diagram of the system and equivalent thermal circuit can be seen in Fig. S15(a). For structures without *h*-BN, we fit $\Delta T(t)$ to Eq. (4) of the main text for each run and extract the TBCs. Each fit only covers data to $t_{\text{fit}}$ where $\Delta T(t_{\text{fit}})$ = 0.25 $\Delta T_0$. We find that 20 independent simulations lead to a converged average TBC for each structure. The time evolution of $\Delta T(t)$, averaged over all runs, and examples of $\Delta T(t)$ using our circuit model with the average TBCs are shown in Figs. S15(b,c).

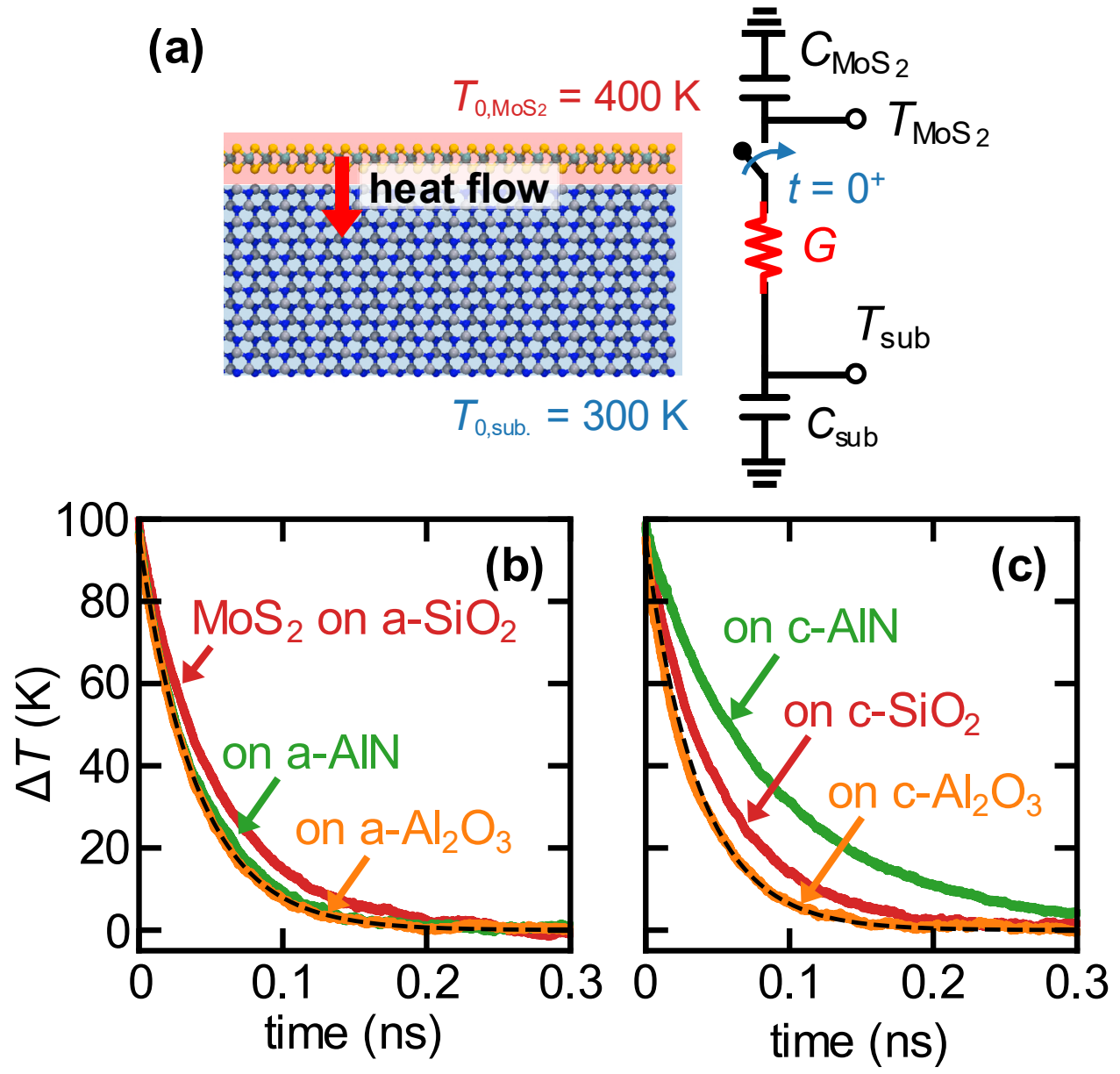

**Figure S15**: **(a)** A schematic of the initial conditions and heat flow direction during an AEMD run of $MoS_2$ on c-AlN. The circuit diagram is the equivalent thermal circuit used to model the cooling process and extract the TBC. **(b)**,**(c)** The average $\Delta T$ of $MoS_2$ on each substrate during the cooling process. The dashed black lines show the output of Eq. (4) of the main text with our fit TBCs.

Since we use Eq. (4) from the main text to fit the temperature profiles and extract the TBC, we must also know the heat capacity of each material. The heat capacity of classic, harmonic solid is $3Nk_B$, where $N$ is the number of atoms and $k_B$ is the Boltzmann constant [31]; however, MD is anharmonic at high temperatures and may deviate from this value [32]. To determine the specific heat, we use the following relation based on energy fluctuations in the NVT ensemble [33]

$$\left\langle (\Delta E)^2 \right\rangle = N k_B T_0^2 c_v . \tag{S4}$$

Here, $T_0$ is the thermostat temperature, $c_v$ is the per-atom heat capacity ($C = N c_v$), $N$ is the number of atoms, and $k_B$ is the Boltzmann constant. We run a 500 ps equilibration in the NVT ensemble, then another 500 ps run in the NVT ensemble where we sample $T$ every 10 fs. From Fig. S16(a), we determine that $c_{v,\text{MoS}_2}/k_B \approx 3.0$, meaning that the heat capacity of our suspended $MoS_2$ matches that of the classical, harmonic value. This is also true of $MoS_2$ supported by any of the substrates. Most substrates also exhibit $c_{v,\text{sub}}/k_B \approx 3$, and, since $C_{\text{sub}} \gg C_{\text{MoS}_2}$ in Eq. (4), the heat capacity term of $MoS_2$ dominates that of the substrate [29] and small deviations do not affect the TBC results. The extraction of heat capacity for c-AlN, using Eq. (S4) is shown in Fig. S16(b).

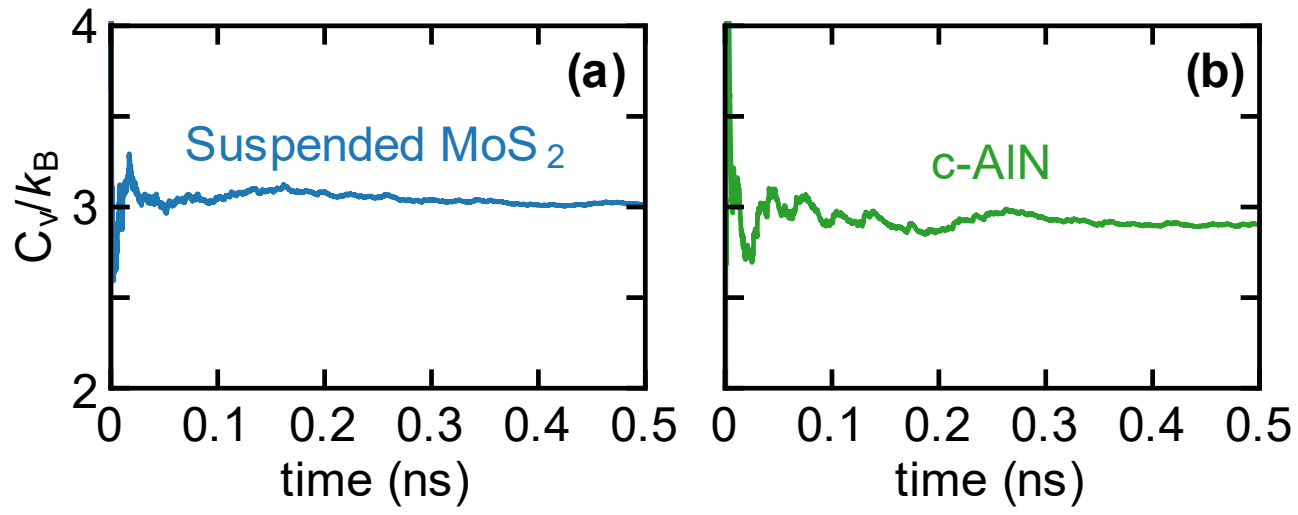

**Figure S16**: The heat capacity of suspended $MoS_2$ **(a)** and c-AlN **(b)** normalized by the Boltzmann constant.

For structures with *h*-BN interlayers (see Fig. S1), the simulation setup stays the same but the equivalent thermal circuit changes. There is a TBC between $MoS_2$ and *h*-BN as well as one between *h*-BN and each substrate. Additionally, *h*-BN has its own heat capacity that must be considered. The equivalent thermal circuit is given in Fig. S17(a) below. We use LTspice IV to simulate the thermal transient, sweeping over a wide range of TBCs between each pair of materials (i.e., $G_{bs}$ for between *h*-BN and a substrate and $G_{mb}$ for between $MoS_2$ and *h*-BN). We compare the temperatures each node [defined in Fig. S17(a)] to the temperatures of each material in an MD simulation. We choose the SPICE simulation with temperature curves that best match the MD simulation and record the TBCs. Here, 'best match' means the minimum mean square error between SPICE and MD temperatures. An example of a best-matched SPICE and MD simulation can be seen in Fig. S17(b).

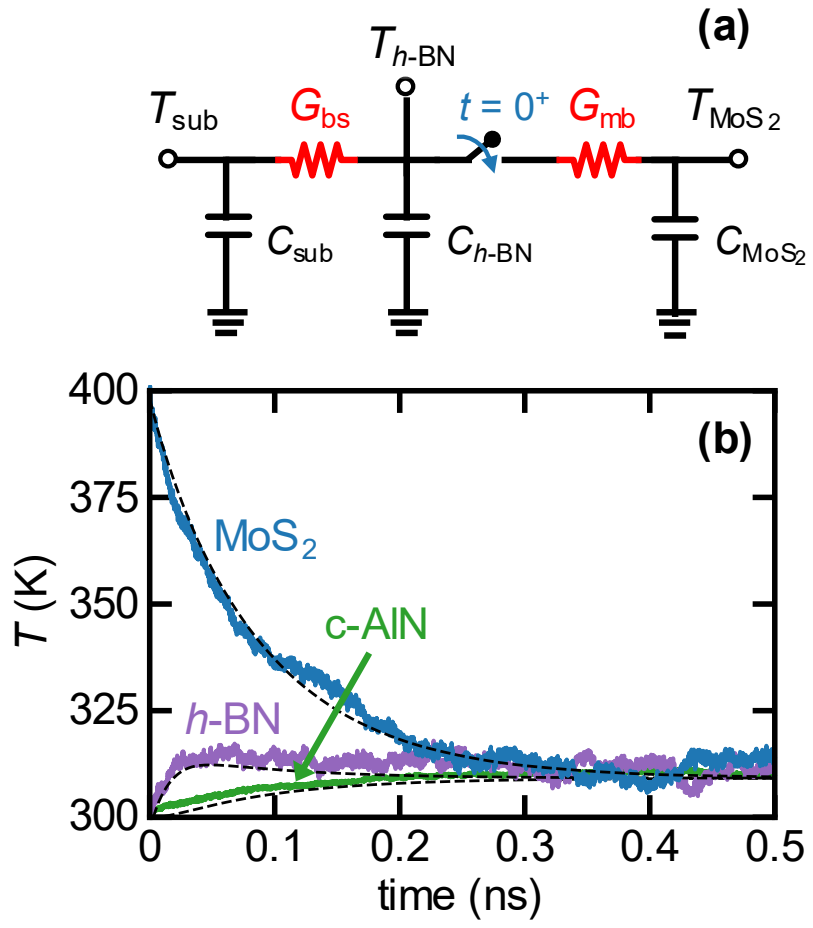

**Figure S17**: **(a)** The equivalent thermal circuit of $MoS_2$ on an *h*-BN-capped substrate with two TBCs to solve for simultaneously. **(b)** Temperatures of $MoS_2$ (blue line), *h*-BN (purple lines), and c-AlN (green lines) in an $MoS_2$/*h*-BN/c-AlN structure. The dashed black lines correspond to the temperatures in the SPICE simulation that best matched the MD temperatures.

Note that the TBCs reported in Table S2 in section S8 for structures with *h*-BN are only for the TBC between $MoS_2$ and *h*-BN [i.e., $G_{mb}$ in Fig. S17(a)]. As evidenced by the small standard error, we find that $G_{mb}$ is stable enough to extract. However, the temperature evolution of *h*-BN is, in general, extremely noisy and often follows the temperature of the substrate closely. As a result, the TBC between *h*-BN and each substrate varies wildly, with averages ranging from ~80 to 340 $MWm^{-2}K^{-1}$. We do not report these values here. Ultimately, in structures with $MoS_2$ supported by *h*-BN-capped substrates, the $MoS_2$/*h*-BN interface limits cross-plane heat removal.

## S8. Thermal Properties Summary

| Structure | $\kappa_{total}$ (Wm$^{-1}$K$^{-1}$) | $\kappa_{in\text{-}plane}$ (Wm$^{-1}$K$^{-1}$) | $\kappa_{out\text{-}of\text{-}plane}$ (Wm$^{-1}$K$^{-1}$) | **TBC** (MWm$^{-2}$K$^{-1}$) |
|---|---|---|---|---|
| $MoS_2$ | 122.05 ± 1.45 | 90.73 ± 1.25 | 31.31 ± 0.84 | N/A |
| $MoS_2$ on a-$SiO_2$ | 37.95 ± 1.52 | 32.77 ± 1.39 | 5.18 ± 0.55 | 22.37 ± 0.44 |
| $MoS_2$ on c-$SiO_2$ | 41.76 ± 1.47 | 35.57 ± 1.37 | 6.19 ± 0.42 | 23.1 ± 0.44 |
| $MoS_2$ on a-AlN | *36.39 ± 1.28* | 30.68 ± 1.19 | 5.72 ± 0.46 | 28.87 ± 0.55 |
| $MoS_2$ on c-AlN | **67.95 ± 1.65** | 53.12 ± 1.38 | 14.83 ± 0.78 | *14.32 ± 0.26* |
| $MoS_2$ on a-$Al_2O_3$ | 38.98 ± 1.28 | 33.47 ± 1.17 | 5.52 ± 0.48 | 31.81 ± 0.63 |
| $MoS_2$ on c-$Al_2O_3$ | 65.75 ± 1.72 | 55.94 ± 1.65 | 9.81 ± 0.67 | **34.3 ± 1.04** |
| $MoS_2$ on *h*-BN/a-$SiO_2$ | 58.19 ± 2.74 | 52.47 ± 2.63 | 5.71 ± 0.75 | 20.23 ± 0.29 |
| $MoS_2$ on *h*-BN/c-$SiO_2$ | 61.04 ± 2.77 | 54.91 ± 2.64 | 6.12 ± 0.79 | 20.07 ± 0.35 |
| $MoS_2$ on *h*-BN/a-AlN | *47.97 ± 2.72* | 43.05 ± 2.57 | 4.92 ± 0.63 | **26.23 ± 0.60** |
| $MoS_2$ on *h*-BN/c-AlN | 71.69 ± 2.83 | 60.27 ± 2.66 | 11.42 ± 0.88 | *18.18 ± 0.3* |
| $MoS_2$ on *h*-BN/a-$Al_2O_3$ | 51.05 ± 3.53 | 46.55 ± 3.47 | 4.5 ± 0.93 | 24.93 ± 0.42 |
| $MoS_2$ on *h*-BN/c-$Al_2O_3$ | **74.47 ± 2.96** | 65.59 ± 2.73 | 8.88 ± 1.02 | 18.96 ± 0.23 |
| $MoS_2$/*h*-BN | 101.22 ± 3.17 | 77.46 ± 2.79 | 23.76 ± 1.13 | N/A |

**Table S2**: The calculated TC and TBC of $MoS_2$ on different substrates. The coloring highlights different substrate materials: $SiO_2$ (red), AlN (green), $Al_2O_3$ (orange), and no substrate (blue). The largest value of a property in each section is bolded and the lowest value is italicized. Note that only the TBC between $MoS_2$ and *h*-BN is listed for structures with an *h*-BN interlayer, and only the TC of $MoS_2$ is being calculated in the $MoS_2$/*h*-BN heterostructure.

## S9. Details of the Analytic Thermal Model

In this section, we present the details of the analytical thermal model. The device geometry is shown in Fig. 6(a) in the main text. Insulating boundary conditions are used for each surface except for the bottom surface of the silicon substrate (not shown), which is set to the ambient temperature. In finite element method (FEM) simulations, the thickness of the silicon substrate is 500 µm, however, the analytical model does not depend on this thickness, and the silicon substrate has minimal influence on the thermal resistance, except for very long channels. The TC of each 50-nm thick insulator film and its TBC with substrate are listed in Table S3. For the contact metal, we assume thermal conductivity $\kappa_{met}$ = 150 Wm$^{-1}$K$^{-1}$ [34] and a metal-$MoS_2$ $TBC_{MoS_2\text{-}met}$ = 20 MWm$^{-2}$K$^{-1}$ [35]. In the analysis presented here, we do not account for the temperature variation of these quantities, as the temperature will depend on the power dissipated in the device. We use the length-dependent thermal conductivity of $MoS_2$ results from the main text, where the length is $L + 2L_{con}$, i.e., the $MoS_2$ underneath the contacts is included in the length for the purpose of thermal conductivity calculation. This also means that, apart from increasing the contact thermal resistance, short contacts also result in a reduced $MoS_2$ thermal conductivity.

| Insulator | $\kappa_{ins}^{\perp}$(Wm$^{-1}$K$^{-1}$) | $\kappa_{ins}^{\parallel}$ (Wm$^{-1}$K$^{-1}$) | TBC with silicon (MWm$^{-2}$K$^{-1}$) |
|---|---|---|---|
| a-$SiO_2$ | 1.4 [36] | 1.4 | 500 [37-40] |
| c-$SiO_2$ | 6.7 [41] | 8.5 | 333 [39,40] |
| a-AlN | 2 [42-44] | 2 | 125 [43-45] |
| c-AlN | 35 [42,46] | 35 | 125 [43-45] |
| a-$Al_2O_3$ | 1.5 [47] | 1.5 | 145 [47] |
| c-$Al_2O_3$ | 18 [46] | 28 | 145 [47] |

**Table S3**: The TC and TBC with silicon of each insulator film. Because the TC of thin films are lower than bulk due to boundary scattering, 50 nm thin film TC values were used where available. $\kappa_{ins}^{\perp}$ indicates the cross-plane (vertical) TC of the insulator and $\kappa_{ins}^{\parallel}$ indicates the in-plane TC. $\kappa_{ins}^{\parallel}$ is only used for the effective channel width calculation. Thus, any unqualified $\kappa_{ins}$ elsewhere in this text refers to $\kappa_{ins}^{\perp}$.

The analytical model of Eq. (6) in the main text is summarized in the form of the following equations.

Thermal healing length for heat spreading in the $MoS_2$ channel:

$$L_{\mathrm{H}} = \sqrt{\frac{W}{R_{\mathrm{MoS_2}}\, g}}$$

Thermal conductance into the substrate per unit length along the channel:

$$g = \left( \frac{1}{W G_{\mathrm{ins}}} + \frac{1}{2\kappa_{\mathrm{sub}}} \sqrt{\frac{L}{W_{\mathrm{eff}}}} \right)^{-1}$$

Contact thermal resistance:

$$R_{\mathrm{con}} = \frac{1}{W} \left[ \begin{array}{l} \sqrt{\dfrac{R_{\mathrm{MoS_2}}}{G_{\mathrm{ins}} + G_{\mathrm{met}}}} \coth\left( \sqrt{R_{\mathrm{MoS_2}} \left( G_{\mathrm{ins}} + G_{\mathrm{met}} \right)} L_{\mathrm{con}} \right) + \\ \left( 1 + \dfrac{G_{\mathrm{ins}}}{G_{\mathrm{met}}} \right)^{-3/2} \sqrt{\dfrac{R_{\mathrm{met}}}{G_{\mathrm{ins}}}} \coth\left( \sqrt{R_{\mathrm{met}} \dfrac{G_{\mathrm{ins}} G_{\mathrm{met}}}{G_{\mathrm{ins}} + G_{\mathrm{met}}}} L_{\mathrm{con}} \right) \end{array} \right]$$

$MoS_2$ in-plane thermal sheet resistance:

$$R_{\mathrm{MoS_2}} = \frac{1}{\kappa t}$$

Contact in-plane thermal sheet resistance:

$$R_{\mathrm{met}} = \frac{1}{\kappa_{\mathrm{met}} t_{\mathrm{con}}}$$

$MoS_2$-to-substrate thermal conductance per unit area:

$$G_{\mathrm{ins}} = \left( \frac{1}{\mathrm{TBC}_{\mathrm{MoS_2\text{-}ins}}} + \frac{W}{W_{\mathrm{eff}}} \frac{1}{\mathrm{TBC}_{\mathrm{ins\text{-}sub}}} + \frac{W}{W_{\mathrm{eff}}} \frac{t_{\mathrm{ins}}}{\kappa_{\mathrm{ins}}} \right)^{-1}$$

$MoS_2$-to-contact metal thermal conductance per unit area:

$$G_{\mathrm{met}} = \mathrm{TBC}_{\mathrm{MoS_2\text{-}met}}$$

Effective width for fringe heat conductance correction:

$$W_{\mathrm{eff}} = W + 2t_{\mathrm{ins}} \frac{1 + \dfrac{\kappa_{\mathrm{ins}}^{\perp}}{t_{\mathrm{ins}} \mathrm{TBC}_{\mathrm{ins\text{-}sub}}}}{\sqrt{12.847 + \dfrac{\kappa_{\mathrm{ins}}^{\perp}}{t_{\mathrm{ins}} \mathrm{TBC}_{\mathrm{ins\text{-}sub}}}}} \sqrt{\frac{\kappa_{\mathrm{ins}}^{\parallel}}{\kappa_{\mathrm{ins}}^{\perp}}}$$

Equation (6) of the main text has been previously used to model self-heating in TMD transistors [48,49] but with a rather simple model for the contact thermal resistance and without including the effects of $\mathrm{TBC}_{\mathrm{MoS2\text{-}met}}$, $\mathrm{TBC}_{\mathrm{ins\text{-}sub}}$, or the contact length $L_{\mathrm{con}}$. In defining the auxiliary quantities $R_{\mathrm{con}}$, $g$, and $W_{\mathrm{eff}}$, we take these quantities into consideration here.

We note that the two terms in $R_{\mathrm{con}}$ are each associated with a mode of heat conduction with a characteristic length (or thermal healing length). The first term has a thermal healing length of $[R_{\mathrm{MoS2}}(G_{\mathrm{ins}} + G_{\mathrm{met}})]^{-1/2}$ and is associated with heat escaping into either the substrate (modeled by $G_{\mathrm{ins}}$) or the contact (modeled by $G_{\mathrm{met}}$) as it flows laterally through $MoS_2$. (Note that the conductances $G_{\mathrm{ins}}$ and $G_{\mathrm{met}}$ are added as they are "in parallel".) The second term has a healing length of $[R_{\mathrm{met}}G_{\mathrm{ins}}G_{\mathrm{met}}/(G_{\mathrm{ins}} + G_{\mathrm{met}})]^{-1/2}$ and is associated with heat spreading laterally in the contact as it leaks into the substrate. (Note that the conductances $G_{\mathrm{ins}}$ and $G_{\mathrm{met}}$ are "in series" as the heat must flow through the contact-$MoS_2$ interface as well as the insulator to reach the substrate.) Because $MoS_2$ is atomically thin and has a moderate thermal conductivity, $R_{\mathrm{MoS2}}$ is quite high and the healing length associated with the first term is very small, usually around $L_{\mathrm{H}} \approx 25$ nm. If the contacts are no shorter than a factor of 1.5 times this healing length, the coth factor can be dropped (i.e., replaced with unity), introducing no more than 10% error and yielding the simplified expression

$$R_{\mathrm{con}} \approx \frac{1}{W}\left[\sqrt{\frac{G_{\mathrm{MoS_2}}}{G_{\mathrm{ins}}+G_{\mathrm{met}}}}+\left(1+\frac{G_{\mathrm{ins}}}{G_{\mathrm{met}}}\right)^{-3/2}\sqrt{\frac{R_{\mathrm{met}}}{G_{\mathrm{ins}}}}\coth\left(\sqrt{R_{\mathrm{met}}\frac{G_{\mathrm{ins}}G_{\mathrm{met}}}{G_{\mathrm{ins}}+G_{\mathrm{met}}}}L_{\mathrm{con}}\right)\right]. \tag{S5}$$

This approximation holds for all materials and geometries considered in this work. The second term depends on the contact length, as longer contacts allow better cooling via heat spreading (unless the contact length is much longer than the associated healing length). For moderately long contacts ($L_{\mathrm{con}} > \sim 500$ nm), or if the contacts provide additional cooling through interconnects (not modeled in this work), the second term is small compared to the first. In this case,

$$\begin{aligned} R_{\mathrm{con}} &\approx \frac{1}{W}\sqrt{\frac{R_{\mathrm{MoS_2}}}{G_{\mathrm{ins}}+G_{\mathrm{met}}}} \\ &= \frac{1}{W}\left\{\kappa t\left[\left(\frac{1}{\mathrm{TBC}_{\mathrm{MoS_2\text{-}ins}}}+\frac{W}{W_{\mathrm{eff}}}\frac{1}{\mathrm{TBC}_{\mathrm{ins\text{-}sub}}}+\frac{W}{W_{\mathrm{eff}}}\frac{t_{\mathrm{ins}}}{\kappa_{\mathrm{ins}}}\right)^{-1}+\mathrm{TBC}_{\mathrm{MoS_2\text{-}met}}\right]\right\}^{-1/2}. \end{aligned} \tag{S6}$$

If $\mathrm{TBC}_{\mathrm{MoS2\text{-}ins}}$ and $\mathrm{TBC}_{\mathrm{MoS2\text{-}met}}$ are much smaller than $\kappa_{\mathrm{ins}}/t_{\mathrm{ins}}$ and $\mathrm{TBC}_{\mathrm{ins\text{-}sub}}$ (a cruder approximation, especially for thermally resistive insulators such as a-$SiO_2$),

$$R_{\mathrm{con}} \approx \frac{1}{W\sqrt{t\kappa\left(\mathrm{TBC}_{\mathrm{MoS_2-ins}}+\mathrm{TBC}_{\mathrm{MoS_2-met}}\right)}}. \tag{S7}$$

To validate the accuracy of the analytical model, we compared its results to those of finite-element method simulations. A comparison is shown in Fig. S18 for four different device geometries. The error in the analytical model is less than 18% for all geometries and materials considered.

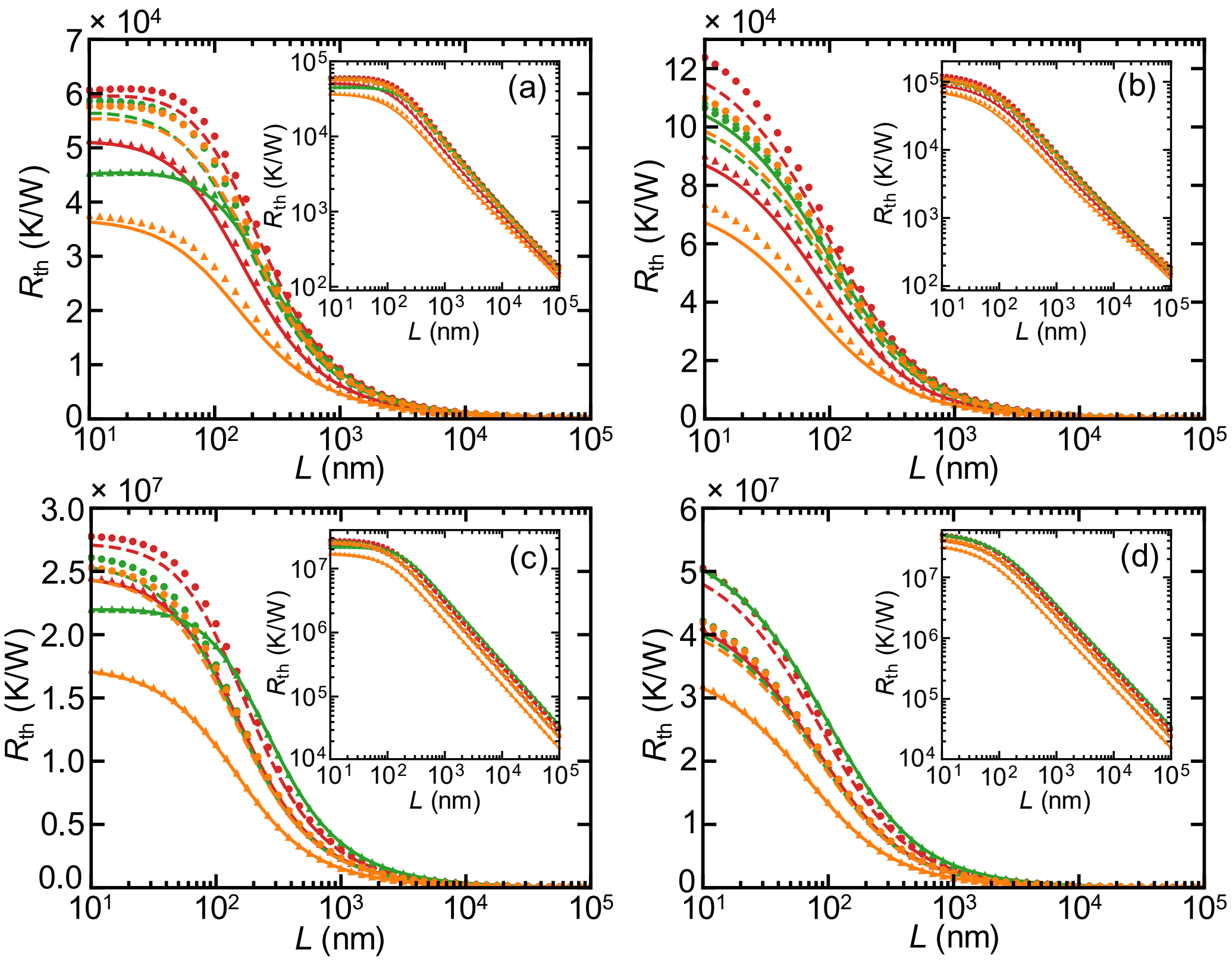

**Figure S18**: Comparison of the analytical model and finite element method simulation results. Red, green and yellow represent $SiO_2$, AlN and $Al_2O_3$, respectively. Solid and dashed correspond to analytical model results for the crystalline and amorphous phases, respectively. Triangles and circles represent simulation results for the crystalline and amorphous phases, respectively. **(a)** Wide (10 µm) transistor with long (10 µm) contacts. **(b)** Wide (10 µm) transistor with narrow (50 nm) contacts. **(c)** Narrow (20 nm) transistor with long (10 µm) contacts. **(d)** Narrow (20 nm) transistor with short (50 nm) contacts.